\documentclass[prd,twocolumn,superscriptaddress,floatfix,amsmath,amssymb,amsfonts,footinbib,longbibliography,showpacs,showkeys]{revtex4-2}
\usepackage{float} 
\usepackage{scalerel}
\usepackage[normalem]{ulem}
\usepackage[english]{babel}
\usepackage{graphicx}
\usepackage{dcolumn}
\usepackage{bm}
\usepackage{blindtext}
\usepackage{verbatim}
\usepackage{relsize}
\usepackage{mathrsfs}
\usepackage{musicography}
\usepackage{amsmath}
\usepackage{blindtext}
\usepackage{cancel}
\usepackage{physics}
\usepackage{epstopdf}
\usepackage{mathtools}
\usepackage{blindtext}
\usepackage{tensor}
\usepackage{color}
\usepackage[usenames,dvipsnames]{pstricks}
\usepackage{epsfig}
\usepackage{pst-grad} 
\usepackage{pst-plot} 
\usepackage{hyperref}
\usepackage{verbatim}
\usepackage{slashed}
\usepackage{dsfont}
\usepackage{amsmath,amssymb,amsthm}  
\usepackage{bbold}
\usepackage{lipsum}
\usepackage{bbm}

\allowdisplaybreaks[1] 

\usepackage{cleveref}

\begin{document}

\title{Interferometrically Enhanced Asymmetry in Strong-field Ionization with Bright Squeezed Vacuum}

\author{G. Singh}
\email{gsingh@perimeterinstitute.ca}

\affiliation{Perimeter Institute for Theoretical Physics, Waterloo, Ontario, N2L 2Y5, Canada}
\affiliation{Department of Physics and Astronomy, University of Waterloo, Waterloo, Ontario, N2L 3G1, Canada}

\author{T. Rook}
\affiliation{Clarendon Laboratory, University of Oxford, Parks Road, Oxford OX1 3PU, United Kingdom}

\author{J. Rivera-Dean}
\affiliation{Department of Physics and Astronomy, University College London, Gower Street, London WC1E 6BT, UK}
\affiliation{
ICFO – Institut de Ciencies Fotoniques, The Barcelona Institute of Science and Technology, 08860 Castelldefels (Barcelona)}

\author{C. Figueira de Morisson Faria}
\affiliation{Department of Physics and Astronomy, University College London, Gower Street, London WC1E 6BT, UK}

\date{\today}
\allowdisplaybreaks
\begin{abstract}
We demonstrate that quantum light statistics can be used to control strong-field ionization at the tunneling step leaving the average driving field unchanged. Using a bichromatic linearly polarized field composed of a strong coherent driver and a weak bright squeezed vacuum (BSV), we show through simulation that the asymmetry of the resulting photoelectron momentum distributions is significantly enhanced, with the skewness increasing by roughly two orders of magnitude compared with an equally weak coherent perturbation. A semiclassical analysis based on the strong-field approximation (SFA) reveals that the effect originates from asymmetric fluctuations in the instantaneous field amplitude, which strongly modify the tunneling ionization probability while leaving the electron’s continuum dynamics essentially unchanged. The resulting amplification provides a proof of concept for enhancing phase-sensitive signatures of sub-cycle ionization dynamics.
\end{abstract}

\maketitle
\allowdisplaybreaks
\section{Introduction}
Symmetry has long served as a fundamental principle in physics and a powerful guide in understanding physical laws. Its role is encapsulated by Noether’s theorem, which establishes the connection between continuous symmetries and conservation laws \cite{Noether1918}. Beyond conservation laws, symmetries provide a unifying framework that systematically constrains dynamics, reduces the effective dimensionality of calculations, determines selection rules, and enables the classification of states and phases, revealing common structures behind otherwise disparate physical phenomena \cite{Yang1954,Gell-Mann1961,Coleman1967,Weinberg1967,Haag1975,Born1927,Jahn1937,Longuet-Higgins1963,Berry1984,Haldane1988}.

In strong-field and attosecond physics, tailored light such as elliptically polarized \cite{Nubbemeyer2008,Abu-samha2011,Hofmann2016,Danek2018}, orthogonal two-color (OTC) \cite{Shafir2012,Das2013,Xie2015,Richter2015,Das2015,Henkel2015,Li2016,Gong2017,Han2017,Zhang2014,Xie2017,Han2018,Tulsky2018}, bicircular \cite{Milos2000,Smirnova2015JPhysB,Milos2015,Milos2016,Mancuso2016,Hoang2017,Almajid2017,Busuladvic2017,Eckart2018,Milos2018,Ayuso2018,Ayuso2018II,Baykusheva2018,Eicke2019,Yue2020,Maxwell2021,Kang2021}, knotted \cite{Pisanty2019} and chiral \cite{Rozen2019,Ayuso2019,Mayer2024} fields, possess spatio-temporal symmetries, which have been systematically studied (for reviews see, e.g., \cite{Habibovic2024,neufeld2025}; for a group-theoretical discussion see \cite{Neufeld2019}). 
Temporal periodicity leads to rings in above-threshold ionization (ATI) photoelectron momentum distributions (PMDs) \cite{Agnostini1979} and half-cycle symmetry (invariance under a translation by half a cycle followed by time-axis reflection) ensures that there are only odd-order harmonics in the high-harmonic generation (HHG) spectrum \cite{Ferray1988}.

Moreover, controlled symmetry-breaking may be employed to probe a system's dynamics. For instance, few-cycle pulses break the temporal translation invariance and thus enable pulse characterization \cite{Paulus2003,Baltuska2003} and isolated attosecond-pulse generation \cite{Hentschel2001,Chen2014}. Instead of using a broad spectral bandwidth, combining two field frequencies breaks the half-cycle symmetry of the light, producing asymmetries in photoelectron emission to characterize tunnel ionization \cite{Shafir2012,Zhao2013,pedatzur_attosecond_2015,Tan2018,Rook2022}. In phase-of-the-phase spectroscopy, the time delay in two-color fields is modulated to map quantum phase differences in photoelectron spectra and structural features in the target \cite{Skruszewicz2015,Almajid2017,Tulsky2018}. 

Often, it is desirable to maximize the effects of spatio-temporal asymmetry in strong-field observables and minimize the disruption to the electron's continuum dynamics.~With that in mind, adding a weak second color to a strong driving field introduces subtle, but measurable distortions in the PMDs. As the strength of the second wave increases, the electron's propagation in the continuum can change significantly \cite{Arbo2015,Luo2017}, leading to  additional ionization pathways and rescattering events,  drastic changes in the dominant ionization events, or even caustics (see \cite{Hashim2025,Weber2026} for examples in an orbit-based SFA framework). This adds complexity to the problem and hinders the ability to reconstruct ionization pathways. This invites the question of whether ionization can be substantially modified, with dramatic consequences for the PMDs, without strongly perturbing the continuum dynamics. 

Quantum light may provide a solution to this question. In particular, Bright Squeezed Vacuum (BSV) can serve as the weak perturbing component of a two-color field, allowing us to investigate whether it helps achieve this goal.~Recent experiments use BSV to drive HHG and strong-field ionization \cite{Spasibko2017,Rasputnyi2024} which has opened a wide range of possibilities for experimenting with shaped light that were previously inaccessible \cite{rivera-dean_structured_2025,lemieux_photon_2025,tzur_measuring_2025,stammer_weak_2025,rivera-dean_attosecond_2025}.~Here, we explore symmetry in ATI for linearly polarized two-color fields, by adding a weaker BSV driver of frequency $\omega$ to an intense coherent field of frequency $2\omega$.~We show that even a weak squeezed second wave influences the resulting PMDs in a radical fashion. Exploiting the field statistics of the weak BSV field yields asymmetric momentum distributions, with a significantly stronger asymmetry than that produced by an equally weak coherent perturbing field. 

To understand the origin of this asymmetry, we extend the symmetry classification of linearly polarized bi- and polychromatic classical fields introduced in Ref.~\cite{Rook2022}. A monochromatic linearly polarized field exhibits three discrete symmetries: (i) half-cycle symmetry, (ii) temporal reflection about the field extrema, and (iii) temporal reflection at the field nodes followed by spatial reflection. Adding a weaker coherent $\omega$-field breaks the half-cycle symmetry and one of the remaining reflection symmetries.~In this work, we investigate how perturbing the field with \textit{quantum} light, specifically BSV light, modifies these symmetries and the resulting PMDs.~This poses an interesting scenario where the average field retains the symmetries of the monochromatic field, while its quantum fluctuations generally do not, therefore providing a distinct mechanism for symmetry breaking in the photoelectron dynamics. 

For this purpose, this work is organized as follows. Section \ref{supp:theory} provides the theoretical framework of the paper, with a coherent-state-based formulation from first principles, a discussion of the key observables and of how our approach relates to other existing derivations, and the saddle-point approximation employed. 
Subsequently, in Sec.~\ref{sec:results}, we discuss the fields and the PMDs, quantify their asymmetry and analyze the physics behind it using specific ionization events and saddle-point solutions. Finally, in Sec.~\ref{sec:conclusions}, we summarize the article and state our main conclusions.

\section{Theoretical framework}
\label{supp:theory}
One of the central observables in ATI is the PMD, which gives the probability of detecting an electron with final momentum $\boldsymbol{p}=(p_x,p_y)$, where $p_x$ and $p_y$ are the electron momentum components parallel and perpendicular to the laser-field polarization. This observable, which is well established from a semiclassical perspective under various levels of approximation~\cite{becker_plateau_2018,faria_it_2020}, encodes interference between electron trajectories ionized at different times within the strong-field cycle, providing insights into both the driving field and the target system. Extensions of the PMD that incorporate the field's Hilbert space, and thus provide a full quantum treatment of the electric field, have been developed previously~\cite{ATIPRL,Lyu2025} using phase-space distributions. Here, we provide a first-principles derivation based instead on a coherent state expansion of the driving field.
\subsection{A coherent-state-based formulation}\label{appx:A1}
Throughout this work, unless otherwise stated, we consider a driving field consisting of a coherent $2\omega$ field and a BSV $\omega$ field, described at the initial time $t_0$ by
\begin{equation}\label{Eq:SM:QO:init}
	\ket{\Phi(t_0)}
		= \big[
				\hat{S}_\omega(\xi)
			\otimes 
				\hat{D}_{2\omega}(\alpha_{2\omega})
			\big]\ket{\bar{0}} 
\end{equation}
where $\hat{D}_\omega(\alpha) = \exp[\alpha_\omega \hat{a}_\omega^\dagger - \alpha^*_\omega \hat{a}_\omega]$ and $\hat{S}_{\omega}(\xi) = \exp[\tfrac12(\xi^*\hat{a}_\omega^2 - \xi \hat{a}_\omega^{\dagger 2})]$ are the displacement and squeezing operators, respectively, acting on a specific optical mode with frequency $\omega$.~Here, $\hat{a}^{(\dagger)}$ denotes the annihilation (creation) operator and $\ket{\bar{0}}$ the vacuum state of all modes. Since the initial state can be expanded in any basis, we are free to choose one that is particularly convenient for the problem at hand. In the context of strong-field physics, the interaction of matter with coherent states of light is well-known~\cite{PRX_Javier,RiveraDean2022}, and therefore the overcomplete coherent state basis is a natural choice, with closure relation $\mathbbm{1} = \pi^{-1} \int \dd^2\alpha  \dyad{\alpha}$ ($\ket{\alpha} \equiv \hat{D}(\alpha)\ket{0}$).~An equivalent representation to Eq.~\eqref{Eq:SM:QO:init} can thus be obtained by inserting this closure relation for the $\omega$ mode in front of the squeezing operator, yielding
\begin{equation}\label{Eq:SM:QO:coh}
	\ket{\Phi(t_0)}
		= \int \dd^2 \alpha_\omega\ c(\alpha_\omega)
				\ket{\alpha_\omega} \otimes \ket{\alpha_{2\omega}},
\end{equation}
where we denote $c(\alpha_\omega) \equiv \pi^{-1}\bra{\alpha_\omega}\hat{S}_\omega(\xi)\ket{0_{\omega}}$, which can be written more explicitly as
\begin{equation}
    \!c(\alpha_\omega)
		\!=\! \dfrac{1}{\pi\sqrt{\cosh(r)}} \!\exp[\!- \dfrac{\abs{\alpha_\omega}^2}{2}
					\!- \!\dfrac12 e^{i\phi} \tanh(r) \alpha_\omega^{*2}].
\end{equation}
In this expression, we have written $\xi = r e^{i\phi}$ where $r\geq 0$ denotes the squeezing strength and $\phi$ determines the phase-space direction along which squeezing occurs.~Here, we restrict to scenarios where this weak $\omega$ field has intensities $I_{\omega} \sim 10^{12}$ W/cm$^2$, while the $2\omega$ coherent state contribution $I_{2\omega} \sim 10^{14}$ W/cm$^2$, corresponding to intensity ratios $I_{\omega}/I_{2\omega} \sim 10^{-2}$.~For the case of squeezed $\omega$ field, such intensity regimes translate to squeezing strengths of up to $r= 12.15$, which is compatible with state-of-the-art BSV generation techniques~\cite{heimerl_multiphoton_2024,Rasputnyi2024,lemieux_photon_2025,tzur_measuring_2025,heimerl_driving_2025} based on high-gain spontaneous-parametric down-conversion~\cite{Spasibko2017,manceau_indefinite-mean_2019}. These high intensities have been theoretically and experimentally shown to strongly affect strong-field processes~\cite{stammer_colloquium_2025}.  

Considering that the atom is initially in its ground state $\ket{\text{g}}$, and denoting by $\hat{U}(t,t_0)$ the unitary operator describing the time-evolution of the system from $t_0$ to $t$, the joint light-matter state at any time $t$ can be written as
\begin{equation}\label{Eq:SM:QO:coh:time:I}
	\ket{\Psi(t)}
		= \!\!\int \!\dd^2 \alpha_\omega \
				c(\alpha_\omega)
					\hat{U}(t,t_0)
						\ket{\text{g}}
						\otimes \ket{\alpha_{\omega}}
						\otimes \ket{\alpha_{2\omega}}.
\end{equation}
In our case, we work under the dipole and single-active-electron approximations, such that $\hat{U}(t,t_0)$ satisfies 
\begin{equation}
	i \pdv{\hat{U}(t,t_0)}{t}
		= \big[
				\hat{H}_{\text{at}}
				+ \hat{\boldsymbol{r}} \cdot \hat{\boldsymbol{E}}
				+ \hat{H}_{\text{field}}
			\big]
				\hat{U}(t,t_0),
\end{equation}
where $\hat{H}_{\text{at}}$ is the atomic Hamiltonian, $\hat{\boldsymbol{r}}\cdot \hat{\boldsymbol{E}}$ denotes the light-matter interaction in the length-gauge, and $\hat{H}_{\text{field}} = \sum_{\omega,\mu} \hbar \omega \hat{a}^\dagger_{\omega,\mu}\hat{a}_{\omega,\mu}$ is the free-field Hamiltonian. The electric field operator is given by $\hat{\boldsymbol{E}} = i \sum_{\omega,\mu}\boldsymbol{\epsilon}_\mu g(\omega) [\hat{a}_{\mu,\omega} - \hat{a}_{\mu,\omega}^\dagger]$ where $\boldsymbol{\epsilon}_\mu$ is the polarization vector, and $g(\omega) = \sqrt{\omega/(2\epsilon_0 V)}$, with $\hbar=1$, is the light-matter coupling parameter.  In this latter expression,  $\epsilon_0$ denotes the vacuum permittivity and $V$ the quantization volume, which for atomic systems satisfies $g(\omega)\sim 10^{-8}$ a.u.~In this work, both optical modes in Eq.~\eqref{Eq:SM:QO:init} are assumed to be linearly polarized along the same direction $\hat{e}_x$, and therefore the reason why we omit the polarization index $\mu$ in the equation and in subsequent ones. Unless otherwise stated, all subsequent  expressions are given in atomic units. 

Within the interaction picture with respect to $\hat{H}_{\text{field}}$, the electric field operator acquires a time-dependence ($\hat{a}_{\mu,\omega} \to \hat{a}_{\mu,\omega}e^{-i\omega t}$, so $\hat{\boldsymbol{E}} \to \hat{\boldsymbol{E}}(t)$), and Eq.~\eqref{Eq:SM:QO:coh:time:I} can be written as
\begin{align}
	\ket{\Psi(t)}
		\!&= \!\!\!\int \!\!\dd^2 \alpha_\omega  c(\alpha_\omega)
				\hat{U}_{\text{f}}(t,t_0)\hat{U}_I(t,t_0)
					\ket{\text{g}}
					\!\otimes\!
					\ket{\alpha_{\omega}}
					\!\otimes\!
					\ket{\alpha_{2\omega}} \nonumber
		\\
		&= \!\!\!\int\!\! \dd^2 \alpha_\omega c(\alpha_\omega)
			\hat{U}_{\text{f}}(t,t_0)
				\big[
					\hat{D}_{\omega}(\alpha_\omega)
						\!\otimes\!
					\hat{D}_{2\omega}(\alpha_{2\omega})
				\big]\nonumber
		\\&\hspace{2.5cm}\times
				\hat{U}_{I,\boldsymbol{\alpha}}(t,t_0)
					\ket{\text{g}}
						\otimes \ket{\bar{0}},
\end{align}
where $i \partial\hat{U}_{\text{f}}(t,t_0)/\partial t = \hat{H}_{\text{field}}\hat{U}_{\text{f}}(t,t_0)$ and $i\partial\hat{U}_I(t,t_0)/t = [\hat{H}_{\text{at}} + \hat{\boldsymbol{r}}\cdot \hat{\boldsymbol{E}}(t)]\hat{U}_I(t,t_0)$.~Furthermore, in going from the first to the second equality we define $\hat{U}_{I,\boldsymbol{\alpha}}(t,t_0) \equiv	[\hat{D}^\dagger_{\omega}(\alpha_\omega)\otimes\hat{D}^\dagger_{2\omega}(\alpha_{2\omega})]\hat{U}_I(t,t_0)[\hat{D}_{\omega}(\alpha_\omega)\otimes\hat{D}_{2\omega}(\alpha_{2\omega})]$, which satisfies
\begin{equation}
	i\pdv{\hat{U}_{I,\boldsymbol{\alpha}}(t,t_0)}{t}
		= \big[
				\hat{H}_{\text{at}}
				+\hat{\boldsymbol{r}} \cdot \boldsymbol{E}_{\boldsymbol{\alpha}}(t)
				+ \hat{\boldsymbol{r}} \cdot \hat{\boldsymbol{E}}
			\big]
			\hat{U}_{I,\boldsymbol{\alpha}}(t,t_0)
\end{equation}
where $\boldsymbol{E}_{\boldsymbol{\alpha}}(t) \equiv \bra{\alpha_{\omega},\alpha_{2\omega},\bar{0}}\hat{\boldsymbol{E}}(t)\ket{\alpha_{\omega},\alpha_{2\omega},\bar{0}}$ is the classical electric field corresponding to the coherent state amplitudes $\boldsymbol{\alpha} \equiv (\alpha_\omega,\alpha_{2\omega})$, more explicitly
\begin{equation}
	\begin{aligned}
	\boldsymbol{E}_{\boldsymbol{\alpha}}(t)
		&= i\boldsymbol{g}(\omega)
			\big[
				\alpha_\omega e^{-i\omega t} - \alpha_\omega^* e^{i\omega t}
			\big]
			\\&\quad
			+ i\boldsymbol{g}(2\omega)
			\big[
				\alpha_{2\omega} e^{-2i\omega t} - \alpha_{2\omega}^* e^{2i\omega t}
			\big].
            \label{eq:Efield}
	\end{aligned}
\end{equation}

Under the SFA~\cite{Lewenstein1994,amini_symphony_2019}, and neglecting the back-reaction of the electronic motion onto the field~\cite{RiveraDean2022,rivera-dean_role_2024}, one can approximate
\begin{equation}\label{Eq:SM:semiclass:sol}
	\begin{aligned}
	&\hat{U}_{I,\boldsymbol{\alpha}}(t,t_0)
		\ket{\text{g}}\otimes \ket{\bar{0}}
		\approx e^{iI_p(t-t_0)}\ket{\text{g}}\otimes \ket{\bar{0}}
			\\& \hspace{0.3cm}
			-i
				\int \dd\boldsymbol{p}
                    \int^{t}_{t_0}
					\dd t'
						b_{\boldsymbol{\alpha}}(\boldsymbol{p},t,t')
						\ket{\boldsymbol{p}+\boldsymbol{A}_{\boldsymbol{\alpha}}(t')}\otimes \ket{\bar{0}},
	\end{aligned}
\end{equation}
where $\boldsymbol{A}_{\boldsymbol{\alpha}}(t) = - \int \dd t \boldsymbol{E}_{\boldsymbol{\alpha}}(t)$ is the vector potential associated with the field, and $b_{\boldsymbol{\alpha}}(\boldsymbol{p},t,t')$ is the direct ATI probability amplitude,
\begin{equation}\label{SM:eq:b_alpha}
	b_{\boldsymbol{\alpha}}(\boldsymbol{p},t,t')
		= e^{i S_{\boldsymbol{\alpha}}(\boldsymbol{p},t,t')}
			\bra{\boldsymbol{p}+\boldsymbol{A}_{\boldsymbol{\alpha}}(t')}
				\hat{\boldsymbol{r}}
			\ket{\text{g}} \cdot \boldsymbol{E}_{\boldsymbol{\alpha}}(t'),
\end{equation}
with $S_{\boldsymbol{\alpha}}(\boldsymbol{p},t,t')$ the semiclassical action given by
\begin{equation}\label{SM:eq:action}
	S_{\boldsymbol{\alpha}}(\boldsymbol{p},t,t')
		= -\dfrac12 \int^{t}_{t'} \dd \tau
				\big[
					\boldsymbol{p}
					+ \boldsymbol{A}_{\boldsymbol{\alpha}}(\tau)
				\big]^2
			+ I_p(t-t').
\end{equation}
Here, $I_p$ is the ionization potential of the atom, i.e., $\hat{H}_{\text{at}}\ket{\text{g}} = -I_p \ket{\text{g}}$. Thus, by combining Eqs.~\eqref{Eq:SM:QO:coh:time:I} and \eqref{Eq:SM:semiclass:sol}, we obtain the final state of the joint light-matter interaction within the assumptions described above, which explicitly reads
    \begin{align}
        \ket{\Psi(t)}
            &= e^{iI_p(t-t_0)} \ket{\text{g}}\otimes \hat{U}_\text{f}(t,t_0)\ket{\Phi(t_0)}\label{eq:11}
            \\& + \int \dd^2 \alpha_\omega\int \dd \boldsymbol{p} \int_{t_0}^t \dd t'
                c(\alpha_\omega)\hat{U}_{\text{f}}(t,t_0)
                    b_{\boldsymbol{\alpha}}(\boldsymbol{p},t,t')
                    \nonumber
                    \\& \times
                    \ket{\boldsymbol{p} + \boldsymbol{A}_{\boldsymbol{\alpha}}(t')}
                    \otimes \ket{\alpha_\omega}
                    \otimes \ket{\alpha_{2\omega}}.    \nonumber
    \end{align}

\subsection{Photoelectron spectrum and relation to other approaches}\label{appxA2}

As stated previously, the PMD corresponds to the probability of measuring photoelectrons, or photoelectron yield, with a given final momentum $\boldsymbol{p}$.~From Eq.~\eqref{Eq:SM:QO:coh:time:I}, this yield can be calculated by projecting onto the final momentum and taking the limit of $t(t_0)$ going to $\infty(-\infty)$: 
\begin{equation}
	\begin{aligned}
	Y(\boldsymbol{p})
			&= \lim_{\substack{t \to \infty\\ t_0\to-\infty }}\tr[
					\big( 
						\dyad{\boldsymbol{p}}
							\otimes \mathbbm{1}
					\big)
					\dyad{\Psi(t)}
					]
			\\&= \lim_{\substack{t \to \infty \\ t_0\to-\infty}}\braket{\Phi(\boldsymbol{p},t)},
	\end{aligned}
\end{equation}
where the identity operator acts on the optical subspace.~In the above expression, $\ket{\Psi(t)}$ is given by Eq.~\eqref{eq:11} with $t_0 \to -\infty$ and we have defined $\ket{\Phi(\boldsymbol{p},t)} = \braket{\boldsymbol{p}}{\Psi(t)}$.~Under the same conditions for which Eq.~\eqref{Eq:SM:semiclass:sol} is valid, this state can be written, up to normalization, as
\begin{equation}\label{Eq:SM:dATI:state}
	\ket{\Phi(\boldsymbol{p},t)}
		\approx \int \!\dd^2 \alpha_\omega\! \int^{t}_{t_0} \dd t'
				c(\alpha)b_{\boldsymbol{\alpha}}(\boldsymbol{p},t,t')\ket{\alpha_{\omega}}\otimes \ket{\alpha_{2\omega}}.
\end{equation}
Accordingly, the ATI yield can be expanded as
\begin{align}\label{Eq:SM:dATI:Spec}
	Y(\boldsymbol{p})
		\approx \!\!\lim_{\substack{t \to \infty \\ t_0 \to -\infty}}\!&\int\! \dd^2 \alpha_\omega\! \int \!\dd^2 \beta_\omega
			\!\int_{t_0}^t \!\!\dd t_1 \int_{t_0}^t \!\!\dd t_2\
			c(\alpha_\omega)c^*(\beta_\omega)\nonumber
			\\& \times
				b_{\boldsymbol{\alpha}}(\boldsymbol{p},t,t_1)
				b^*_{\boldsymbol{\beta}}(\boldsymbol{p},t,t_2)
				\braket{\beta_\omega}{\alpha_{\omega}}.
\end{align}

One of the main differences between the above expression and previous approaches (see, e.g., Refs.~\cite{Wang2023,Lyu2025,stammer_colloquium_2025} and references therein) is that it includes off-diagonal elements of the quantum-optical density matrix, i.e., terms with $\beta_\omega \neq \alpha_\omega$, and at first sight appears to lack a Husimi distribution weighting of the semiclassical ATI photoelectron probabilities.~These differences arise from the fact that in many of those approaches, the so-called \emph{classical limit} is employed~\cite{Gorlach2023}, where both $V$ and $\alpha_{\omega}$ are taken to infinity under the condition that the ratio $\varepsilon =2g(\omega)\alpha_\omega \propto \alpha_\omega/\sqrt{V}$, which determines the electric field strength, remains constant.~In contrast, in our analysis, we keep $g(\omega)$ fixed and small, satisfying $g(\omega)\sim 10^{-8} \ll 1$~\cite{RiveraDean2022,rivera-dean_role_2024}. In the following, we therefore justify under which conditions retaining only the diagonal contributions of Eq.~\eqref{Eq:SM:dATI:Spec} provides an accurate approximation.

To this end, note that our integral contains two systems of different nature whose behavior varies with $\alpha$:~the electron, through $b_{\alpha}(\boldsymbol{p},t,t_1)$, and the light, through the rest of the integrand.~The relevant off-diagonal terms are weighted by the coherent state overlap  $\braket{\beta}{\alpha} \propto \text{exp}[-\abs{\beta-\alpha}^2/2]$, which is exponentially localized around $\alpha = \beta$.~Since $\alpha$ and $\beta$ parametrize the classical field amplitude associated with each mode via $E_\alpha = 2g(\omega) \alpha$, a difference $\abs{\beta-\alpha}$ corresponds directly to a field difference $\delta E = 2g(\omega)\abs{\beta-\alpha}$.~Thus, terms with $\abs{\alpha-\beta}\gtrsim 5$, strongly suppressed by a factor of $10^{-6}$, correspond to field differences of only $\delta E \simeq 10^{-7}$ a.u., which is perturbative with respect to the driving field strength, of order $10^{-2}$ a.u.~A perturbative expansion of $b_{\alpha}(\boldsymbol{p},t,t_1)$ around $\alpha$ then gives (see Appendix~\ref{Sec:SM:ATI:prob:amp:exp})
\begin{equation}
    \begin{aligned}
    &b_{\alpha+\delta\alpha}(\boldsymbol{p},t,t_1)
            \approx b_{\alpha}(\boldsymbol{p},t,t_1)
    \\&\hspace{0.2cm}       
         \times 
         \Bigg[1
                +\delta E_{\alpha}
                \Bigg(
                    \dfrac{1}{E_{\omega}+E_{2\omega,\alpha}} + \dfrac{\sqrt{U_p}}{\omega^2}
                        +\dfrac{1}{\omega\sqrt{I_p}}
                \Bigg)
        \Bigg],
    \end{aligned}
\end{equation}
which, for the parameters used in this work, $\delta E_\alpha\sim 10^{-7}$ a.u. yields corrections of order $10^{-5}-10^{-6}$ on $b_{\alpha}(\boldsymbol{p},t,t_1)$. Consequently, over most of the support of $\braket{\beta}{\alpha}$ one has $b_{\alpha}(\boldsymbol{p},t,t_1)\simeq b_{\beta}(\boldsymbol{p},t,t_1)$, which upon substitution in Eq.~\eqref{Eq:SM:dATI:Spec} results in
\begin{equation}
    \begin{aligned}\label{SM:eq:P_pf_full}
   Y(\boldsymbol{p})
		&\approx \int \dd^2 \alpha_\omega Y_{\boldsymbol{\alpha}}(\boldsymbol{p})
            \int \dd^2\beta_\omega \
            c(\alpha_\omega) c^*(\beta_\omega) \braket{\beta_\omega}{\alpha_\omega}
        \\&= \int \dd^2 \alpha_\omega \ Q(\alpha_\omega)
			     Y_{\boldsymbol{\alpha}}(\boldsymbol{p}),
    \end{aligned}
\end{equation}
where $Y_{\boldsymbol{\alpha}}(\boldsymbol{p}) =\lim_{\substack{t \to \infty \\ t_0 \to -\infty}} \vert\int_{t_0}^t \dd t' b_{\boldsymbol{\alpha}}(\boldsymbol{p},t,t')\vert^2$ is the semiclassical ATI probability for a field with coherent state amplitudes determined by $\boldsymbol{\alpha}$ (see Eq.~\ref{SM:eq:b_alpha}), evaluated in this work using the saddle-point approximation~(see Sec.~\ref{Sec:SPA}), and $\pi\abs{c(\alpha_\omega)}^2 = \pi^{-1}\vert\bra{\alpha_\omega}\hat{S}_\omega(\xi)\ket{0}\vert^2\equiv Q(\alpha_\omega)$, i.e., the Husimi distribution of the BSV, given by
\begin{equation}\label{SM:eq:BSV_Husimi}
    Q_{\text{BSV}}(\alpha;r,\phi)\!=\!\frac{1}{\pi \cosh r}e^{-|\alpha|^2-\frac{\tanh r}{2}\Big(e^{-i\phi}\alpha^2 + e^{i\phi}\alpha^{*2}\Big)}.    
\end{equation}
If the temporal limits are not taken in $Y_{\boldsymbol{\alpha}}(\boldsymbol{p})$, one obtains the differential ionization probability, $Y_D(\mathbf{p},t, t_0)$ for an electron ionized within the finite time interval $[t_0,t]$.

Note that for $\phi=0$, $Q_{\text{BSV}}$ is squeezed (anti-squeezed) along $\mathrm{Re}(\alpha)$ ($\mathrm{Im}(\alpha)$) and positive increment in $\phi$ rotates the Husimi distribution counter-clockwise.~For comparison, we also consider coherent and thermal fields~\footnote{Although throughout this work we have considered pure states as initial driver states, any mixed state can be purified by enlarging the Hilbert space: in particular, a thermal state can be obtained by tracing out one mode of a two-mode squeezed state. In our setting, the auxiliary mode would not participate in the strong-field interaction with the atom and can, for instance, be physically separated into a different spatial mode.} in our analysis; their Husimi functions are given by
\begin{align}
    Q_{\text{coherent}}(\alpha;\alpha_0) &= \frac{1}{\pi}\exp{-|\alpha-\alpha_0|^2},\\
    Q_{\text{thermal}}(\alpha;\bar{n})&=\frac{1}{\pi(1+\bar{n})}\exp{-\frac{|\alpha|^2}{1+\bar{n}}},
\end{align}
where $\alpha_0$ is the amount of displacement in the complex $\alpha$ plane and $\bar{n}$ is the average number of photons in the thermal state.~For the weaker $\omega$ field, $\alpha_0$ is taken to be one-tenth of the electric field amplitude, consistent with its intensity being $1/100$ of that of the stronger field ($E_0 \propto \sqrt{I_0}$).~Accordingly, $\bar{n}$ is chosen to be $1/100$ of the average photon number in the $2\omega$ field.

As observed, the yield $Y(\boldsymbol{p})$ in Eq.~\eqref{SM:eq:P_pf_full} corresponds to an average over semiclassical PMDs evaluated for electric fields of the form in Eq.~\eqref{eq:Efield}, in agreement with the literature~\cite{Wang2023,Lyu2025,stammer_colloquium_2025}, where the field amplitude of the $\omega$ field, $\boldsymbol{\alpha}=(\alpha_\omega, \alpha_{2\omega})$ where $\alpha_\omega$ is sampled from the corresponding Husimi distribution. In our case, and without loss of generality, we take $\alpha_{2\omega}$ to be real, so that the relative phase between the two colors is controlled only by the phase of $\alpha_{\omega}$. For bichromatic fields with a weaker coherent or thermal component, varying this phase amounts to varying the temporal phase $\theta$, i.e., the time delay between the two waves. In contrast, for the bichromatic BSV field, we vary the phase of $\alpha_\omega$ by tuning the squeezing angle $\phi$ of the $\omega$ mode, while keeping $\theta$ fixed. A shift $\Delta \phi$ in the squeezing angle corresponds to $\Delta \theta=\Delta\phi/2$ in the temporal phase [Appendix \ref{supp:PMDonr}]. In what follows, $\theta=\pi/2$ ($\theta=\pi/4$) for the bichromatic coherent or thermal light corresponds to the same relative phase as $\phi=0$ ($\phi=-\pi/2$) for the bichromatic BSV field.

\subsection{Saddle point approach to modeling ionization}\label{Sec:SPA}
We solve $Y_{\boldsymbol{\alpha}}(\boldsymbol{p})$ by saddle point approximation for the time integral. This involves finding the stationary point for the argument of the exponential in Eq.~\eqref{SM:eq:b_alpha}
\begin{equation}\label{SM:eq:SPA}
    \partial_{t'} S\big|_{t_{sp}} = 0 \Rightarrow 
	\dfrac{[\boldsymbol{p} + \boldsymbol{A}_{\boldsymbol{\alpha}}(t_{sp})]^2}{2}
		+	I_p = 0,
\end{equation}
which is the saddle point equation. We solve for the (complex) saddle point solutions $t_{sp}$ numerically for each value of $\alpha$ sampled from the Husimi distribution at some value of $r$ and $\phi$. $Y_{\boldsymbol{\alpha}}(\boldsymbol{p})$ after the saddle-point approximation is given by 
\begin{align}\label{SM:eq:P_alpha_SPA}
    Y_{\boldsymbol{\alpha}}(\boldsymbol{p}) &= \Big|\sum_{t_{sp}} \sqrt{\frac{2\pi i}{\partial_{t}^2 S_{\boldsymbol{\alpha}}(\boldsymbol{p},t_{sp})}}e^{i S_{\boldsymbol{\alpha}}(\boldsymbol{p},t_{sp})}\nonumber\\
			&\bra{\boldsymbol{p}+\boldsymbol{A}_{\boldsymbol{\alpha}}(t_{sp})}
				\hat{\boldsymbol{r}}
			\ket{\text{g}} \cdot \boldsymbol{E}_{\boldsymbol{\alpha}}(t_{sp})\Big|^2,
\end{align}
where $\partial_{t}^2 S_{\boldsymbol{\alpha}}(\boldsymbol{p},t_{sp}) = \boldsymbol{E}_{\boldsymbol{\alpha}}(t_{sp})\cdot(\boldsymbol{p}+\boldsymbol{A}_{\boldsymbol{\alpha}}(t_{sp}))$. The saddle-point times and their variation under sampling of $\alpha$ from the Husimi distribution are discussed in the next subsection. To avoid ambiguities associated with the temporal window, we perform incoherent sums in the PMDs as described in Ref.~\cite{Werby2021}. 

We consider the ground state wavefunction of the electron to be given by
$$\bra{\boldsymbol{r}}\ket{g} = \sqrt{\frac{\lambda^3}{\pi}}e^{-\lambda r}\,,$$
where $\lambda$ is the effective nuclear charge felt by the electron and $r$ is the radial parameter. The dipole transition matrix element $\bra{\boldsymbol{p}+\boldsymbol{A}_{\boldsymbol{\alpha}}(t)}\hat{\boldsymbol{r}}\ket{\text{g}} $ is given by \cite{podolsky_momentum_1929}
\begin{equation}
\bra{\boldsymbol{p}+\boldsymbol{A}_{\boldsymbol{\alpha}}(t)}\hat{\boldsymbol{r}}\ket{\text{g}} \!=\! i\sqrt{\frac{\lambda^3}{2}}\frac{16\lambda}{\pi}\frac{-(\boldsymbol{p}+\boldsymbol{A}_{\boldsymbol{\alpha}}(t))}{(\lambda^2\!+\!p_{\perp}^2\!+\!(p_{\parallel}+A_{\boldsymbol{\alpha}}(t))^2)^3}\,.   
\end{equation}
For the 1s state of the Helium atom, $I_p=0.904$ a.u.~and $\lambda=1.6875$. 


\section{Results}
\label{sec:results}
\subsection{Photoelectron momentum distributions}
\begin{figure*}
    \centering
    \includegraphics[width=0.89\linewidth]{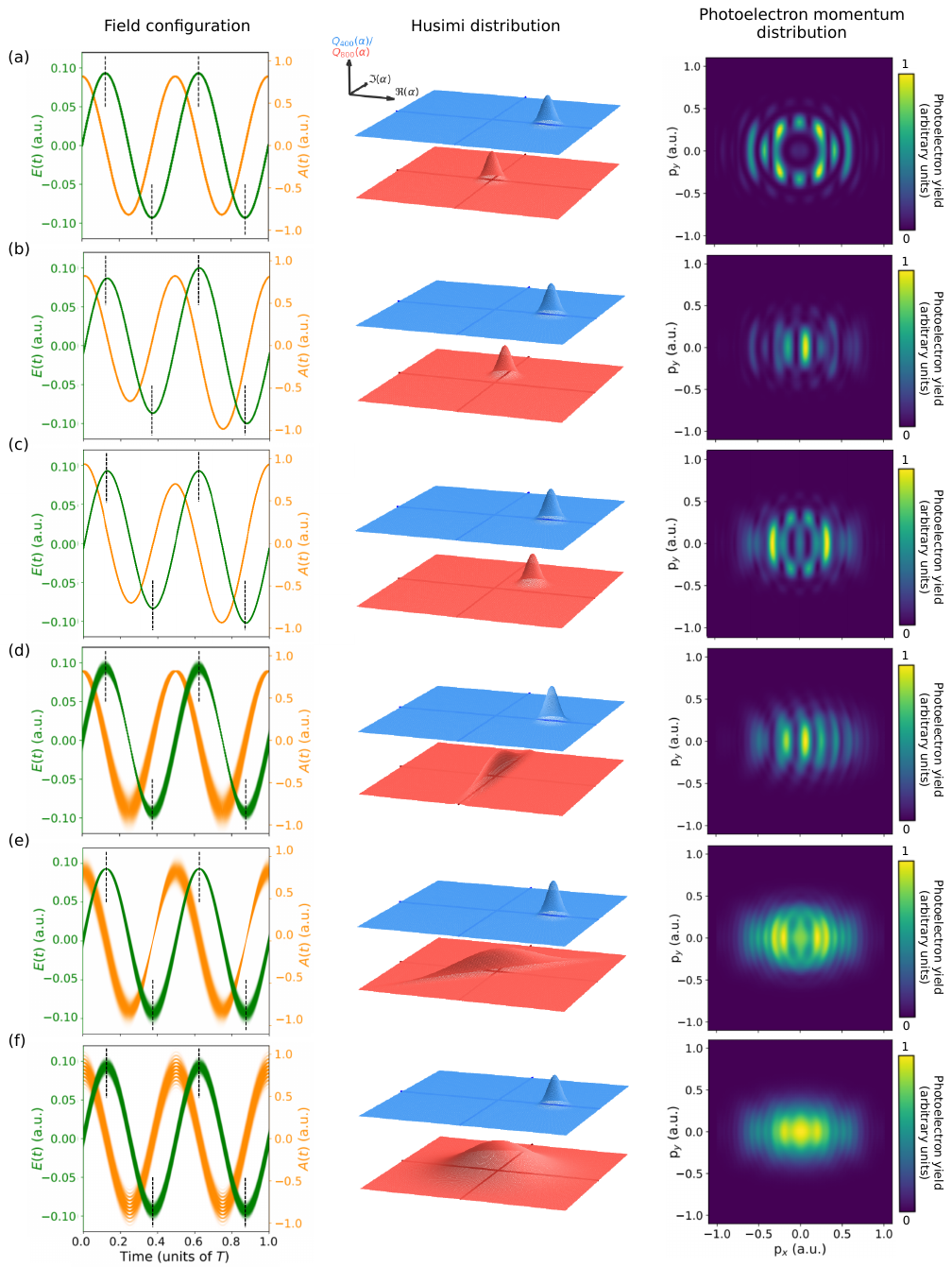}
\caption{Electric fields, Husimi distributions and direct ATI photoelectron momentum distributions (left, middle and right columns, respectively) for He atom ($I_p=0.904$) in linearly polarized coherent light with intensity $I_{\text{coh}} = 3\times 10^{14}$ W/cm$^2$ and wavelength $\lambda = 400$ nm where: (a) is unperturbed monochromatic light and the remaining rows are perturbed by a $I=3\times 10^{12}$ W/cm$^2$, wavelength $\lambda = 800$ nm, (b) coherent field with relative phase $\theta = \pi/2$, (c) coherent field with relative phase $\theta=\pi/4$, (d) and (e) squeezed vacuum, with squeezing parameter $r=12.15$ corresponding to the given intensity, and squeezing angle (d) $\phi = 0$ and (e) $\phi = -\pi/2$, and finally (f) thermal light. The PMDs are calculated using $Y((p_x,p_y))$ [Eq.~\eqref{SM:eq:P_pf_full}]. The dashed lines in the left column mark the field extrema, and the time is in units of the period $T=2\pi/\omega$ of the $\omega$ field. The PMDs are normalized by the maximum value in each panel. The photoelectron yield corresponding to the bichromatic BSV light is roughly $10\%$ higher than the bichromatic coherent light.}
    \label{fig:Fig1}
\end{figure*}

Fig.~\ref{fig:Fig1} illustrates the electric fields (and the associated vector potentials) employed in this work, together with the corresponding Husimi distributions for the two colors and the resulting PMDs (left, middle and right columns, respectively) for He atom. 
For reference, the first row shows these quantities for a monochromatic, linearly polarized field, while the remaining rows correspond to bichromatic fields defined in Eq.~\eqref{eq:Efield}, where a weaker (b,c) coherent, (d,e) squeezed, or (f) thermal $\omega$ field is superposed onto a stronger coherent $2\omega$ field.
The PMD obtained for the monochromatic, purely coherent wave [Fig.~\ref{fig:Fig1}(a)] is reflection-symmetric with regard to the $p_y$ axis, exhibits clear ATI rings (concentric structures corresponding to discrete photon absorption) and well-defined intra-cycle interference patterns arising from electron trajectories released within the same optical cycle. Adding a coherent wave of frequency $\omega$ [Figs.~\ref{fig:Fig1}(b) and (c)] slightly disrupts these patterns, but retains the sharp interference fringes. 
In contrast, adding the $\omega$-BSV field [Figs.~\ref{fig:Fig1}(d) and (e)] dramatically changes the PMDs. The ATI rings are significantly suppressed, the interference patterns are blurred and the ATI cutoff energy is extended. Both features are due to the uncertainty introduced by the squeezed light: the blurring is associated with loss of coherence, and the cutoff extension is associated with the uncertainty in intensity. 
Similar features are present for thermal light [Fig.~\ref{fig:Fig1}(f)] and have been reported elsewhere for monochromatic squeezed fields~\cite{Lyu2025}.  

Furthermore, for $\phi=0$ [Fig.~\ref{fig:Fig1}(d)], the $p_x \rightarrow -p_x$ symmetry no longer holds. In contrast, for $\phi=-\pi/2$ [Fig.~\ref{fig:Fig1} (e)], the distributions retain this symmetry. A direct comparison with bichromatic waves in which, instead, a coherent second wave of frequency $\omega$ was taken [Fig.~\ref{fig:Fig1}(b), for $(\theta=\pi/2)$ ] shows that the asymmetry in the PMDs is markedly reduced, while choosing a second coherent wave with time delay $\theta=\pi/4$ keeps the $p_x \rightarrow -p_x$ symmetry in the PMD. This raises the question of what physical effect is behind this striking asymmetry amplification. 

The symmetry of the driving field provides insight into this behavior: while, for the monochromatic wave, the symmetries (i)-(iii) are present, the fields in Figs.~\ref{fig:Fig1}(b) and (d)[Figs.~\ref{fig:Fig1}(c) and (e)] retain only the (iii) [(ii)] symmetry. However, the field asymmetry is much more noticeable in the BSV case than in its classical counterparts. For a thermal wave [Fig.~\ref{fig:Fig1}(f)] the asymmetry in the field and in the PMD is absent. In fact, varying the relative phase of the thermal light has no effect on the PMDs due to the rotational symmetry of the thermal state in phase space. The PMDs for intermediate squeezing parameters are given in Appendix \ref{supp:PMDonr}. 

\subsection{Asymmetry characterization}
\begin{figure}
    \centering
    \includegraphics[width=\linewidth]{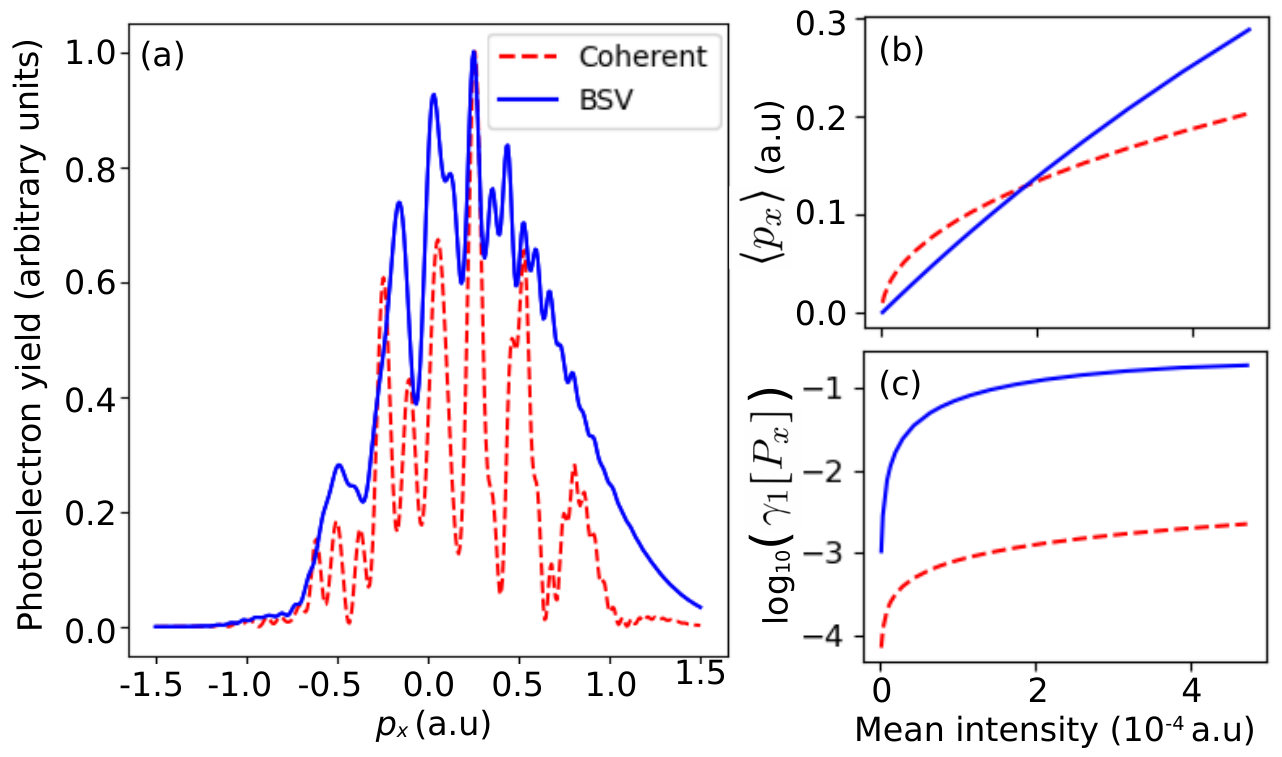}
    \caption{ Panel (a): Dependence of photoelectron yield $Y((p_x,0))$ at the parallel momentum axis on $p_x$, for the coherent bichromatic field [same as Fig.~\ref{fig:Fig1}(b)] and BSV bichromatic field [same as Fig.~\ref{fig:Fig1}(d)] indicated by dashed-red lines and solid-blue lines, respectively. Panels (b) and (c): The asymmetries of the differential ionization probability, $Y_D((p_x,0),t, t_0)$ where the interval $[t_0,t]$ contains only a single ionization peak, for the two types of field studied in panel (a) are characterized by $\langle p_x\rangle$ and the skewness, $\gamma_1 [P_x]$, in terms of their dependence on the mean intensity of the perturbing 800 nm field. This means, we only consider photoelectrons produced within a single half-cycle of the coherent $2\omega$ field with $p_y=0$ and avoid oscillations due to interference between photoelectrons emitted in different half-cycles.}
    \label{fig:asymmetry}
\end{figure}

The asymmetry trend present in Fig.~\ref{fig:Fig1} is confirmed by Fig.~\ref{fig:asymmetry}, where we show that the central moments of the momentum distribution in $p_x$ related to asymmetry, tend to be much larger if an $\omega$-BSV wave at $\phi=0$ is considered. Fig.~\ref{fig:asymmetry}(a) compares the photoelectron momentum distributions along the $p_x$ axis ($p_y=0$), taking the $\omega$ wave to be coherent or squeezed for the asymmetric cases [Figs.~\ref{fig:Fig1}(b) and \ref{fig:Fig1}(d)]. It shows an increase in the asymmetry of the PMDs and a blurring of the peaks for the BSV case and $p_x>0$. More systematically, upon introduction of the $\omega$ perturbing field, the symmetry of the momentum distribution can respond in two qualitatively distinct ways. Firstly, the whole distribution may be shifted; the degree to which this occurs is mainly characterized by the mean value of $p_x$ as a function of mean intensity of the weak $\omega$-field [shown in Fig.~\ref{fig:asymmetry}(b)]. The coherent-bichromatic field approximately exhibits $\langle p_x\rangle~\propto~\sqrt{\langle I_\omega\rangle}$ while the BSV-bichromatic field maintains a roughly linear dependence on mean intensity in the intensity range considered. 
Additionally, the distribution may be skewed [characterized by the skewness of $p_x$ shown in Fig.~\ref{fig:asymmetry}(c)], meaning that the tail of the distribution is more prominent in one direction than another. 
For the normalised probability distribution slice  
\begin{equation}
P_x(p_x)=\frac{Y_D\bigl((p_x,0),t,t_0\bigr)}
    { \int dp'_x\,Y_D\bigl((p'_x,0),t,t_0\bigr)} ,
    \end{equation}
 one can define the skewness as its third standardized moment
    \begin{equation}
    \gamma_1[P_x]=\frac{\displaystyle \int dp_x\,(p_x-\langle p_x\rangle)^3 P_x(p_x)}{\left[\displaystyle \int dp_x\,(p_x-\langle p_x\rangle)^2 P_x(p_x)\right]^{3/2}}.
    \end{equation}

The BSV-bichromatic field leads to a momentum distribution which is orders of magnitude more skewed than for the coherent-bichromatic field. 

\begin{figure*}
    \centering
    \includegraphics[width=0.58\linewidth]{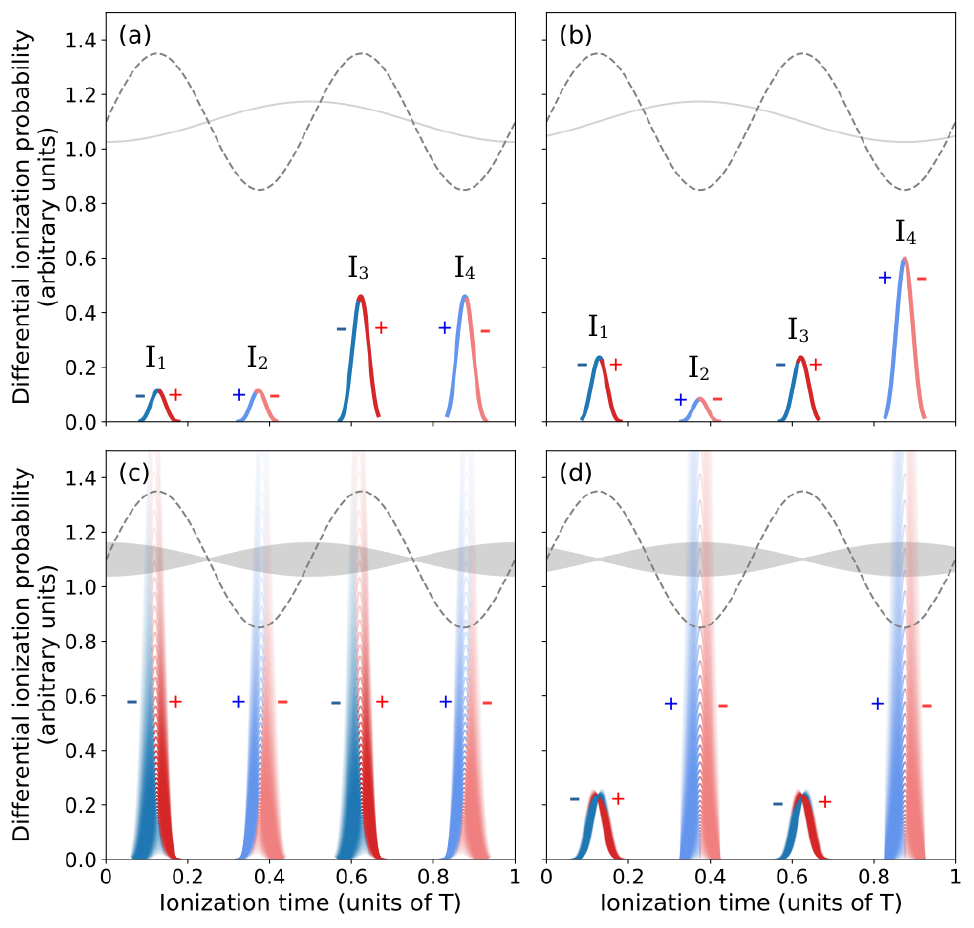}
    \caption{ The dependence of differential ionization probability $Y_D((p_x,0),t, t_0)$  upon the ionization time, computed for temporal windows $[t_0, t]$ defined in the first, second, third and fourth quarter cycle of the $\omega$ field,  is shown for a coherent-bichromatic field with (a) relative (temporal) phase $\theta=\pi/2$ [same as Fig.~\ref{fig:Fig1}(b)] and (b) $\theta=\pi/4$ [same as Fig.~\ref{fig:Fig1}(c)], a BSV-bichromatic field with (c) $\phi=0$ [same as Fig.~\ref{fig:Fig1}(d)] and (d) $\phi=-\pi/2$ [same as Fig.~\ref{fig:Fig1}(e)]. Notably, the relative phase between the two colors in panels (a) and (b) is identical to that in panels (c) and (d), respectively. The times are plotted in units of the period of the $\omega$ field, which is $T=2\pi/\omega$. Whether $p_x$ of the ionized photoelectron is positive/negative is indicated by $+/-$. The short orbits, for which the electron leaves \textit{after} the field extrema, are indicated by red lines and the long orbits, for which the electron is freed \textit{before} the field extrema, are indicated by blue lines, and different tones of red and blue have been chosen for different half cycles. The transparency of each line indicates the magnitude of the Husimi-Q distribution. In each panel, a schematic of the $\omega$ (light gray solid line) and $2\omega$ (dark gray dotted line) fields is shown for reference. The ionization events are labelled in the top-row panels for each half cycle of the $\omega$ field.}
    \label{fig:placeholder4}
\end{figure*}

\subsection{Effect on the electron's ionization events}
For the direct-ATI SFA, the photoelectrons after ionization follow the driving electromagnetic field. In Appendix \ref{supp:statisticsforce}, we show that 
the net force associated with the light statistics of the squeezed component has negligible effect on the propagation of photoelectrons in the continuum. Consequently, changes to the PMDs depend dominantly on the ionization probability. To explore this, in Fig.~\ref{fig:placeholder4} we plot the differential ionization probability associated with specific ionization events (I$_1$--I$_4$), for the bichromatic fields in Figs.~\ref{fig:Fig1}(b), (c), (d) and (e). 

For the coherent fields, in Figs.~\ref{fig:placeholder4}(a) and (b), the ionization times are well defined around the maximum, and the ionization probabilities are sharp. This sharpness leads to well-defined interference fringes in the PMDs displayed in Figs.~\ref{fig:Fig1}(a) and (b). Fig.~\ref{fig:placeholder4}(a), for $\theta=\pi/2$, lacks the half-cycle symmetry and the reflection symmetry (ii) around the extrema. For this figure, the ionization probabilities I$_3$, I$_4$ associated with events in the second half cycle are dominant. Furthermore, all curves associated with I$_1$-I$_4$ lack reflection symmetry about the field extrema, but are reflection-symmetric about its crossing. This lack of symmetry slightly skews the momenta, as observed in Fig.~\ref{fig:Fig1}(b), although the effect is very subtle. For the scenario displayed in Fig.~\ref{fig:placeholder4}(b) ($\theta=\pi/4$), there is only one dominant event, namely I$_4$. The ionization probabilities of the remaining events are orders of magnitude smaller, and, in practice,  will not contribute. One should also note that there is reflection symmetry about the extrema for the probabilities related to $I_2$ and $I_4$, but not for $I_1$ and $I_3$, although the latter two curves are the mirror images of each other. This reflection symmetry, together with a single dominant event, explains why the PMD in Fig.~\ref{fig:Fig1}(c) is essentially reflection-symmetric about $p_x=0$, although the driving field lacks half-cycle symmetry. 

The BSV driver introduces several modifications. First, the noise introduced by the BSV perturbing field [solid grey schematic curves in Fig.~\ref{fig:placeholder4}(c) and (d)] is periodic with half the temporal period compared to a weak coherent perturbation [solid grey schematic curves in Fig.~\ref{fig:placeholder4}(a) and (b)]. This results in ionization patterns which repeat twice as frequently in the BSV-bichromatic case, increasing the number of dominant events. Specifically, for a BSV-bichromatic field at $\phi=0$ [Fig.~\ref{fig:placeholder4}(c)], the zero crossings of both constituent waves are aligned. Furthermore, all events are equally dominant, with the corresponding ionization probability curves repeating every half cycle and being reflection-symmetric about the field zero crossings. The reflection  symmetry about the field extrema, however, is broken, which makes the ionization probability  consistently biased towards positive momenta [notice the dominance of the positive momenta part of the curves at high ionization probability in Fig.~\ref{fig:placeholder4}(c)]. One should note that, in contrast to what would happen to a half-cycle symmetric field, each field contribution within the noise keeps the original periodicity. Hence, the ionization events being half-cycle symmetric do not guarantee that a PMD will be reflection-symmetric about $p_x=0$. This supports the pronounced PMD asymmetry in Fig.\ref{fig:Fig1}(d).

For a BSV-bichromatic field at $\phi=-\pi/2$, the extrema are aligned and there is one dominant event per half-cycle. Events $I_2$ and $I_4$ prevail, and the reflection symmetry (ii) about the extrema holds. Consequently, the left-right asymmetry in the ionization probability curves and in the PMDs disappears as expected [see Fig.\ref{fig:Fig1}(e)]. 

We observe that the BSV contribution can significantly enhance ionization probability in general [compare Fig.~\ref{fig:placeholder4}(a) and (b) with Figs.~\ref{fig:placeholder4}(c) and (d)], since, despite the mean intensity being the same in each case, there is some chance of a measurement of the BSV having a much larger electric field than in the coherent case. This leads to a disproportionate contribution to the tunneling probability, which depends exponentially on the instantaneous field amplitude. The fuzziness observed in the PMDs with a BSV driver [Figs.~\ref{fig:Fig1}(d) and (e)] can be attributed to the uncertainty in the corresponding ionization probabilities, and, for $\phi=-\pi/2$, to their unequal value in subsequent half cycles. The latter weakens the contrast of interference fringes in the PMD [compare Fig.~\ref{fig:Fig1}(d) and (e) with reference to Fig.~\ref{fig:placeholder4}(c) and (d), respectively, and their coherent counterparts]. 

\subsection{Analysis of saddle-point ionization times}
\label{supp:saddlepoints}
\begin{figure*}
    \centering
    \includegraphics[width=0.6\linewidth]{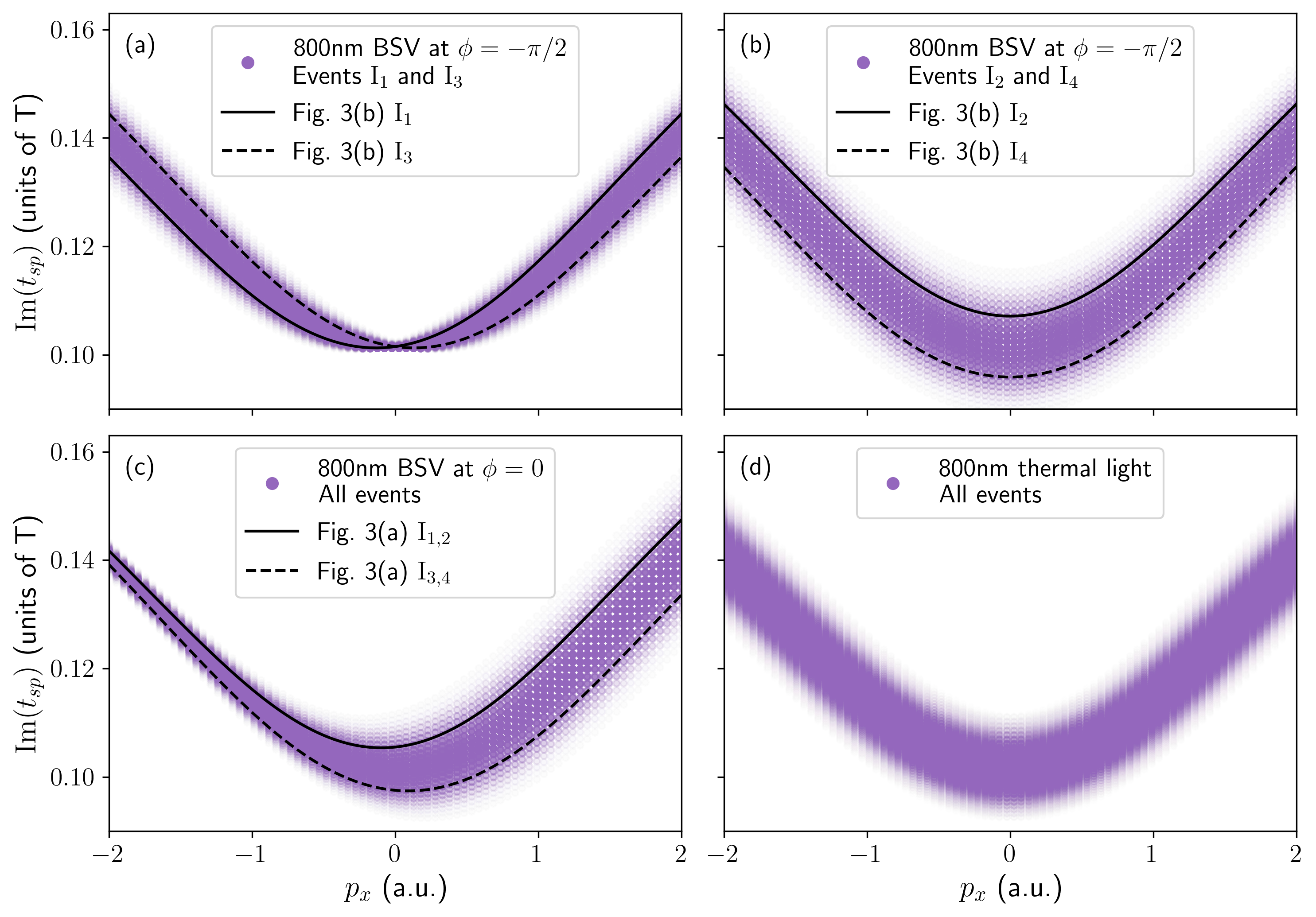}
    \caption{Dependence of imaginary parts of the saddle-point solutions on $p_x$ for linearly polarized bichromatic field with intensity $3 \times 10^{14}$ W/cm$^2$ for the 400 nm coherent part of the light and (a) superposed 800 nm BSV field at $r=12.15$ ($I_{\text{BSV}}= 3\times 10^{12}$ W/cm$^2$) with $\phi=-\pi/2$ for ionization events $I_1$ and $I_3$ as shown in Fig.~\ref{fig:placeholder4} and for (b) events $I_2$ and $I_4$; (c) superposed 800nm BSV field at $\phi=0$ for all events and (d) 800 nm thermal light for all events. The black dotted and dashed lines depict the case of coherent bichromatic field for the labelled events in Fig.~\ref{fig:placeholder4}(a) and \ref{fig:placeholder4}(b).} 
    \label{fig:Sp_im_times}
\end{figure*}
Further details on the dependence of tunneling times on light statistics are provided below. We analyze the saddle point ionization times (obtained from solving Eq.~\eqref{SM:eq:SPA}) as a function of the longitudinal momentum $p_x$ and the complex amplitudes $\alpha$ as sampled from the Husimi distribution of different driving fields. 

From Eq.~\eqref{SM:eq:SPA}, it can be seen that the saddle point solution $t_{sp}=t_r+i t_i$ depends on the momentum $p_x$ and complex variable $\alpha$. For each value of $p_x$ and $\alpha$, we obtain the saddle point solution and plot the imaginary part of the ionization time in Fig.~\ref{fig:Sp_im_times}. The solutions are weighted by the Husimi function through their transparency. The imaginary parts $\mathrm{Im}[t_{sp}]$ provide qualitative information about the instantaneous tunneling probability at $t_{sp}$: the smaller $\mathrm{Im}[t_{sp}]$, the higher the probability for the electron to tunnel through the barrier. This is a consequence of this probability being proportional to $\exp[-2\mathrm{Im}(S)]$ and the dominant term in the action being linear with $t_{sp}$. This explains why the minimum of the imaginary part occurs near $p_x=0$, corresponding to the peak of the field. 

For all phases, the figure shows that introducing squeezing leads to an uncertainty in $\mathrm{Im}[t_{sp}]$, in comparison to the results obtained for coherent bichromatic field (see the black solid and dashed lines in the figure). In principle, all solutions within the shaded regions will contribute to the ATI signal, which explains the stronger ionization probabilities in the BSV case. The corresponding $\mathrm{Im}(t_{sp})$ for the ionization events associated with the coherent bichromatic field with relative temporal phase $\theta=\pi/2$ [Fig.~\ref{fig:placeholder4}(a) and Fig.~\ref{fig:Sp_im_times}(c)] and $\theta=\pi/4$ [Fig.~\ref{fig:placeholder4}(b) and Fig.~\ref{fig:Sp_im_times}(a,b)] are shown as black lines and act as reference markers for the noisy imaginary-time solutions arising from the BSV field.

For $\phi=-\pi/2$ [Figs.~\ref{fig:Sp_im_times}(a) and (b)], the imaginary parts are reflection-symmetric with regard to $p_x=0$, for the BSV case. This means that the tunneling probability and, consequently, the PMD, are reflection-symmetric about $p_x \rightarrow -p_x$. For the classical fields, the scenario is dependent on the event in question. The imaginary parts of the saddle-point solutions associated with $I_1$, $I_3$ [see black lines in Fig.~\ref{fig:Sp_im_times}(a)], for instance, are not symmetric with regard to $p_x=0$, but are the mirror image of each other. However, as shown in Figs.~\ref{fig:placeholder4}(b) and (d), ionization associated with these events is strongly suppressed. 

The remaining imaginary times for $\phi=-\pi/2$ are given in 
Fig.~\ref{fig:Sp_im_times}(b), and correspond to the ionization events $I_2$ and $I_4$ in Fig.~\ref{fig:placeholder4}(d). For the BSV, $\mathrm{Im}[t_{sp}]$ has a larger degree of uncertainty, extending towards lower values. This will result in a larger probability amplitude, in comparison to the contributions from panel (a) which depicts ionization events $I_1$ and $I_3$ in Fig.~\ref{fig:placeholder4}(d) and will worsen the contrast in the quantum-interference patterns, leading to blurred PMDs [see Fig.~\ref{fig:Fig1}(e) in comparison with Fig.~\ref{fig:Fig1}(c)]. Note as well that, for classical fields, a smaller $\mathrm{Im}[t_{sp}]$ indicates that $I_4$ is dominant [see dashed black line in Fig.~\ref{fig:Sp_im_times}(b)], as confirmed by the associated ionization probability in Fig.~\ref{fig:placeholder4}(b).

For $\phi=0$ [Fig.~\ref{fig:Sp_im_times}(c)], $\mathrm{Im}[t_{sp}]$ is no longer reflection-symmetric about $p_x=0$ and exhibits a higher degree of uncertainty for positive momentum $p_x$, reaching much lower values. This will result in a higher signal for positive momenta, as shown in the PMDs. An inspection of Fig.~\ref{fig:placeholder4}(c) confirms this trend, with comparable, but asymmetric ionization probabilities around each driving-field extrema. For a classical field, there will be an asymmetry, but only the events $I_3$, $I_4$ in the second half-cycle will dominate. This is indicated by the smaller $\mathrm{Im}[t_{sp}]$ for these events [dashed black line in Fig.~\ref{fig:Sp_im_times}(c)].
Finally, in Fig.~\ref{fig:Sp_im_times}(d) we show $\mathrm{Im}[t_{sp}]$ for the thermal light. The plot is symmetric about $p_x=0$ and the uncertainty is uniform across the momenta, which explains the blurred and symmetric PMD in Fig.~\ref{fig:Fig1}(f).

We observe that the spread in the imaginary part of the ionization time, for different values of $\alpha$ sampled from the Husimi distribution, varies with $p_x$. This spread can be quantified by the Husimi-weighted variance of the imaginary part of the saddle-point times. For each longitudinal momentum $p_x$, this is defined as
\begin{equation}\label{SM:eq:Variance}
\mathrm{Var}_Q\!\big[\operatorname{Im}(t_{sp})\big]\!\!=\!\frac{ \int \dd^2\alpha\, Q(\alpha)\big(\operatorname{Im}(t_{sp}(\alpha)) - \langle \operatorname{Im}(t_{sp})\rangle_Q\big)^2}
{\int \dd^2 \alpha\, Q(\alpha)},
\end{equation}
where the weighted mean is given by
\begin{equation}
\langle \operatorname{Im}(t_{sp})\rangle_Q
=
\frac{\int \dd^2 \alpha\, Q(\alpha)\,\operatorname{Im}(t_{sp}(\alpha))}
{\int \dd^2 \alpha\, Q(\alpha)}.
\end{equation}

\begin{figure}
    \centering
    \includegraphics[width=0.74\linewidth]{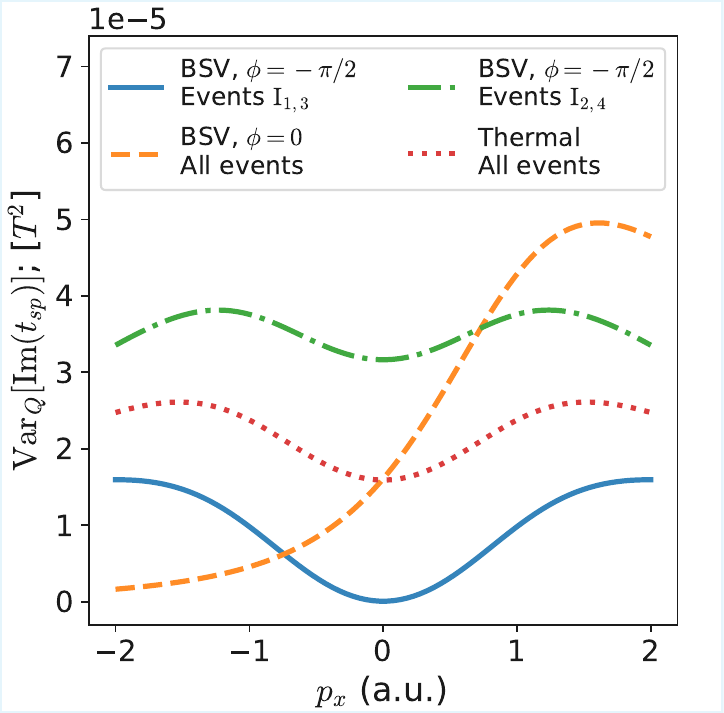}
    \caption{Variance of imaginary part of saddle point times weighted by the Husimi distributions corresponding to the four bichromatic field configurations shown in Fig.~\ref{fig:Sp_im_times}, as a function of $p_x$.}
    \label{fig:Sp_im_times_variances}
\end{figure}

Here, $t_{sp}(\alpha)$ denotes the saddle-point time corresponding to a field realization labeled by $\alpha$, sampled from the Husimi distribution $Q(\alpha)$. The weights $Q(\alpha)$ encode the probability of each phase-space configuration, ensuring that the variance captures the uncertainty in tunneling-time induced by quantum fluctuations of the field in phase space.

Fig.~\ref{fig:Sp_im_times_variances} shows the variance in imaginary part of saddle point times calculated using Eq.~\eqref{SM:eq:Variance} for the plots shown in Fig.~\ref{fig:Sp_im_times}. This provides an additional signature of the asymmetry in the PMDs. For the bichromatic BSV field at $\phi=0$, the variance is asymmetric about $p_x=0$, whereas for $\phi=-\pi/2$ it is symmetric, consistent with the behavior of the PMD and the differential ionization probabilities. We further observe that the variance in the imaginary-time solutions associated with the ionization events $\mathrm{I}_2$ and $\mathrm{I}_4$ is larger than that for $\mathrm{I}_1$ and $\mathrm{I}_3$, in agreement with the trends seen in Fig.~\ref{fig:Sp_im_times}(a,b). Finally, for bichromatic thermal light, the variance of $\mathrm{Im}(t_{sp})$ remains symmetric about $p_x=0$.
\section{Summary and Conclusions}
\label{sec:conclusions}
In summary, the present results show that 
adding a weak bright squeezed vacuum (BSV) field of frequency $\omega$ to a strong coherent wave of frequency $2\omega$ may significantly amplify asymmetries in the resulting ATI photoelectron momentum distributions, in comparison with classical, coherent or thermal fields. These asymmetries occur for values of the relative phase between the two waves, where the reflection symmetry regarding the field extrema breaks down. They can be traced back to the tunneling probabilities associated with each ionization event, which are unequally influenced by the field statistics. Systematic comparison between the skewness of PMDs generated by squeezed and coherent light, respectively, as well as a direct inspection of the ionization probability for specific events has confirmed that this effect is caused by the BSV light statistics and increases with the squeezing parameter. 

Further studies, presented in the Appendix, have shown that, for the parameter range considered in this article, there is no effective photon statistic force altering the electron's propagation in the continuum. Thus, the effect reported here is exclusively associated with the tunneling dynamics, mapping sub-cycle emission times onto measurable asymmetries in photoelectron spectra. 

For the field-traced observable considered here, namely the PMD, and in the small light-matter coupling regime relevant to the present work, the quantum result can be represented by a Husimi-weighted ensemble of coherent-state responses \cite{Gorlach2023,rivera-dean_structured_2025}. Thus, the observed PMD asymmetry should not be interpreted as a standalone witness of nonclassicality, while retaining the physically correct description of the incident field as BSV quantum light. 
This observable-level equivalence, however, does not imply that either the incident BSV or the joint light-matter state is classical. Rather, it results from tracing over the field degrees of freedom together with the separation of field-strength scales associated with small $g(\omega)$. The coherences discarded in the PMD remain present in the joint light-matter state and can become accessible through conditional measurements. Indeed, conditioning on the detected photoelectron has recently been shown to project the driving field into non-Gaussian states that cannot be reproduced by a classical mixture with the same Husimi function \cite{RiveraDean2026}.

Although the present results are reported for direct ATI, we expect the main conclusions to remain relevant for rescattered electrons, since the underlying mechanism originates from tunnel ionization. Moreover, rescattered ATI spectra are typically orders of magnitude weaker than the direct contribution, including for squeezed drivers \cite{Wang2023,Habib2026}. Therefore, the dominant features of the total PMD in the momentum region considered here should remain governed by direct electrons. This expectation is also consistent with our previous TDSE and Coulomb distorted CQSFA calculations for classical fields with a weak secondary driver, where the symmetry changes were found to arise predominantly from tunnelling \cite{Rook2022}. Nevertheless, continuum propagation may introduce additional effects associated with the field statistics in rescattered ATI, whose investigation lies beyond the scope of the present work.

In principle, the asymmetry amplification reported in this work could enable the extraction of tunneling times, quantum phases, and pathway-resolved dynamics. This is in stark contrast with strong-field tunneling with classical fields, for which the relevant information, such as timing, phase, or electron pathway selection, is often encoded in very small differences buried under symmetric backgrounds. Although the present work focuses on a proof-of-concept statement, rather than a concrete road map for experiments,  it demonstrates that one can amplify asymmetry and create new classes of dominant events in some cases solely by altering the tunnelling dynamics. Thus, the effect triggered by a weak second wave became much more prominent, and in principle easier to measure.

This calls for intricate schemes such as the phase-of-the-phase spectroscopy, or observables like the asymmetry parameter, to extract this information.  A weak second harmonic has been employed in, for instance, phase-of-the-phase spectroscopy \cite{Skruszewicz2015}, in which a linearly polarized two-color field is used to tag emission times, and uses changes in ionization probability as a function of the relative phase between two waves. These changes are mapped to the final PMD in such a way that the trajectory with the largest change is dominant. A similar scheme would be possible with the BSV fields used here, with the squeezing parameter being an additional knob and the changes being amplified. Here, we have demonstrated that even a weak additional BSV driver increases the asymmetry by roughly two orders of magnitude and turns a weak differential signal into a robust observable. 
\vspace{-0.5cm}
\section*{Acknowledgements}
We profited from discussions with M. Hassan, H. Ma, P. H. Bucksbaum, D. B. Milo\v sevi\'c, P. Stammer and M. Lewenstein. This work is funded by the UK Engineering and Physical Sciences Research Council (EPSRC) Funding, Grant UKRI2300 - Attosecond Photoelectron Imaging with Quantum Light (APIQuL).


%

\appendix
\section*{Appendix}
 Here, we present material of technical or complementary nature, which supports the findings discussed in the main body of the paper. This includes a proof that for the parameter range employed there is no effective continuum force arising from the field statistics (Appendix \ref{supp:statisticsforce}), PMD dependence on squeezing parameters and linking the time delay with the squeezing parameter (Appendix \ref{supp:PMDonr}), and a justification for 
 the scaling provided when expanding $b_{\alpha + \delta\alpha}(\boldsymbol{p},t,t_1)$ around $\alpha$ (Appendix \ref{Sec:SM:ATI:prob:amp:exp}).

\section{About the photon statistics force}
\label{supp:statisticsforce}

With this background in mind, we can now obtain some practical insight into the role of a photon statistics force. In Refs.~\cite{even_tzur_photon-statistics_2023,rivera-dean_structured_2025}, the influence of squeezing on electron trajectories leading to strong-field phenomena---specifically high-harmonic generation---was interpreted as the action of an effective force arising from the photon statistics of BSV light: the so-called \emph{photon statistics force}.~Its derivation is based on evaluating the Fourier transform of the expectation value of the dipole operator under the saddle-point approximation with respect to all integration variables, including those defining the electric field amplitude.~Yet, this quantity determines the HHG response~\cite{Lewenstein1994,amini_symphony_2019}; because ATI spectra are not determined by dipole expectation values, that derivation cannot be directly transplanted to describe the influence of a photon-statistics force in ATI processes.

In standard semiclassical ATI theory, the saddle-point approximation is instead applied to the probability amplitudes themselves.~Accordingly, in our case we introduce the saddle-point approximation directly at the level of Eq.~\eqref{Eq:SM:dATI:state}. To do so, we first expand the coherent state $\ket{\alpha_\omega}$ in the Fock basis so that all dependence on $\alpha$ and $t'$ appears explicitly in the probability amplitudes~\cite{rivera-dean_role_2024}, i.e.,
\begin{equation}
	\begin{aligned}
	\ket{\Phi(\boldsymbol{p},t)}
		&= \sum_{n=0}^{\infty} \int \dd^2 \alpha \int^t_{t_0} \dd t'
				c(\alpha)
				b_{\boldsymbol{\alpha}}(\boldsymbol{p},t,t') 
					e^{-\frac12\abs{\alpha}^2}
					\\&\hspace{3cm}\times
						\dfrac{\alpha^n}{\sqrt{n!}} \ket{n_{\omega}}\otimes \ket{\alpha_{2\omega}}.
	\end{aligned}
\end{equation}
Note that we renamed the integration variable $\alpha_\omega$ to $\alpha$. From the above equation, we can extract an effective quantum optical action
\begin{equation}
	\begin{aligned}
	&S^{(\text{QO})}_{\boldsymbol{\alpha}}(\boldsymbol{p},t,t')
		= - \dfrac12 
			\int^{t}_{t'} \dd \tau 
				\big[
					\boldsymbol{p} + \boldsymbol{A}_{\boldsymbol{\alpha}}(\tau)
				\big]^2
			+ I_p (t'-t_0)
			\\&\hspace{1cm}
			+i
			 \bigg[
					\dfrac{\alpha_x^2}{e^{-r}\cosh(r)}
					+ \dfrac{\alpha_y^2}{e^{r}\cosh(r)}
					+ \dfrac{\alpha_x^2}{2}
					+ \dfrac{\alpha_y^2}{2}
				\bigg]
			\\&\hspace{1cm}
				+ \alpha_x \alpha_y\tanh(r),
	\end{aligned}
\end{equation}
consisting of the semiclassical action in Eq.~\eqref{SM:eq:action} and additional terms coming from $c(\alpha)$ and the Fock state expansion.

Analogously to Refs.~\cite{even_tzur_photon-statistics_2023,rivera-dean_structured_2025}, we apply the saddle-point approximation to the integration variables $(\alpha_x,\alpha_y,t')$, which leads to the following set of saddle-point equations
\begin{align}
	&\partial_{t'} S^{(\text{QO})}_{\boldsymbol{\alpha}} = 0 \Rightarrow 
	\dfrac{[\boldsymbol{p} + \boldsymbol{A}_{\boldsymbol{\alpha}}(t')]^2}{2}
		+	I_p = 0, \label{Eq:SM:SP:ion:QO}
	\\&\partial_{\alpha_x} S^{(\text{QO})}_{\boldsymbol{\alpha}} = 0 \Rightarrow 
		-\dfrac{g(\omega)}{\omega}
			\int^{t}_{t'} \dd \tau
				\big[ 
					\boldsymbol{p} + \boldsymbol{A}_{\boldsymbol{\alpha}}(\tau)
				\big]\cdot \boldsymbol{\epsilon}_{\mu}
					\cos(\omega \tau)\nonumber
			\\&\hspace{1cm}	
				+ i \alpha_x
					\bigg[
						\dfrac{2}{e^{-r}\cosh(r)}
						+ 1
					\bigg]  \label{Eq:SM:SP:alphax}
				+ \alpha_y\tanh(r) = 0,
	\\&\partial_{\alpha_y} S^{(\text{QO})}_{\boldsymbol{\alpha}} = 0 \Rightarrow 
	-\dfrac{g(\omega)}{\omega}
			\int^{t}_{t'} \dd \tau
				\big[ 
					\boldsymbol{p} + \boldsymbol{A}_{\boldsymbol{\alpha}}(\tau)
				\big]\cdot \boldsymbol{\epsilon}_{\mu}\sin(\omega \tau)\nonumber
		\\&\hspace{1cm}	
			+ i \alpha_y
			\bigg[
				\dfrac{2}{e^{r}\cosh(r)}
				+ 1
			\bigg] \label{Eq:SM:SP:alphay}
	+ \alpha_x \tanh(r) = 0,
\end{align}
where we have taken into account that
\begin{equation}
	\!\!\boldsymbol{A}_{\boldsymbol{\alpha}}(t)
		\!=\! 	\boldsymbol{A}_{2 \omega }(t)
			+ 
				\boldsymbol{\epsilon}_{\mu}
					\dfrac{2g(\omega)}{\omega}
				\big[
					\alpha_x \cos(\omega t)
					+	\alpha_y \sin(\omega t)
				\big].
\end{equation}
Eq.~\eqref{Eq:SM:SP:ion:QO} is identical to that in Section~\ref{Sec:SPA} and gives energy conservation upon ionization, while the remaining equations are associated with the photon statistics force. 

To gain insight into how these expressions scale with $g(\omega)$, we use the fact that the intensity of a BSV state is given by
\begin{equation}
	I_{\text{squ}} = g(\omega)^2 \sinh[2](r).
\end{equation}
By inverting this relation, we obtain
\begin{equation}
	r = \sinh[-1](\pm \dfrac{\sqrt{I_{\text{squ}}}}{g(\omega)}),
\end{equation}
where we have restricted ourselves to real-valued squeezing amplitudes. Using the additional relations
\begin{equation}
	\begin{aligned}
	&e^{\pm \sinh[-1](x)} = \sqrt{x^2 +1} \pm x,
	\\&
	\cosh(\sinh[-1](x)) = \sqrt{x^2+1},
	\end{aligned}
\end{equation}
we find
\begin{align}
	&e^{-r}\cosh(r)
	= \dfrac{1}{g(\omega)^2}
	\Big[
	\sqrt{I_{\text{squ}} + g(\omega)^2}
	- \sqrt{I_{\text{squ}}}
	\Big]\nonumber
	\\&\hspace{2cm}\times
	\sqrt{I_{\text{squ}} + g(\omega)^2},
	\\&e^{r}\cosh(r)
	= \dfrac{1}{g(\omega)^2}
	\Big[
	\sqrt{I_{\text{squ}} + g(\omega)^2}
	+ \sqrt{I_{\text{squ}}}
	\Big]\nonumber
	\\&\hspace{2cm}\times
	\sqrt{I_{\text{squ}} + g(\omega)^2}.
\end{align}
In the strong squeezing regime, we have that $I_{\text{squ}} \gg g(\omega)^2$.~We may therefore consider a Taylor expansion of the square root terms around this limit
\begin{equation}
	\sqrt{1 + \dfrac{g(\omega)^2}{I_{\text{squ}}}}
	\simeq 1 + \dfrac{g(\omega)^2}{2 I_\text{squ}}, 
\end{equation}
which yields
\begin{align}
	&e^{-r}\cosh(r)
	\simeq \dfrac12 + \dfrac{g(\omega)^2}{8I_{\text{squ}}}
	\approx \dfrac12,
	\quad e^{r}\cosh(r)
	\simeq \dfrac{2I_{\text{squ}}}{g(\omega)^2},
\end{align}
where only the dominant higher-order from each Taylor expansion have been retained.~Under these approximations, Eqs.~\eqref{Eq:SM:SP:alphax} and \eqref{Eq:SM:SP:alphay} reduce to
\begin{align}
	&-\dfrac{g(\omega)}{\omega}
		\int^{t}_{t'}\dd \tau	
			\bigg\{
				\big[
					\boldsymbol{p}(\tau)
					+ \boldsymbol{A}_{\boldsymbol{\alpha}}(\tau)
				\big]\cdot
				\boldsymbol{\epsilon}_\mu
				\cos(\omega \tau)
			\bigg\}\nonumber
			\\&\hspace{2cm}
			+ i5 \alpha_x
			+\alpha_y = 0,\label{Eq:SM:SP:alphax:II}
	\\&	-\dfrac{g(\omega)}{\omega}
		\int^{t}_{t'}\dd \tau
			\bigg\{
				\big[
					\boldsymbol{p}(\tau)
					+ \boldsymbol{A}_{\boldsymbol{\alpha}}(\tau)
				\big]\cdot
				\boldsymbol{\epsilon}_\mu
				\sin(\omega \tau)
			\big]
		\bigg\}
			\nonumber
			\\&\hspace{2cm}
			+ i \dfrac{2g(\omega)^2\alpha_y}{I_{\text{squ}}}
			+ i \alpha_y
			+ \alpha_x = 0\label{Eq:SM:SP:alphay:II}.
\end{align}

\begin{figure}
    \centering
    \includegraphics[width=1\columnwidth]{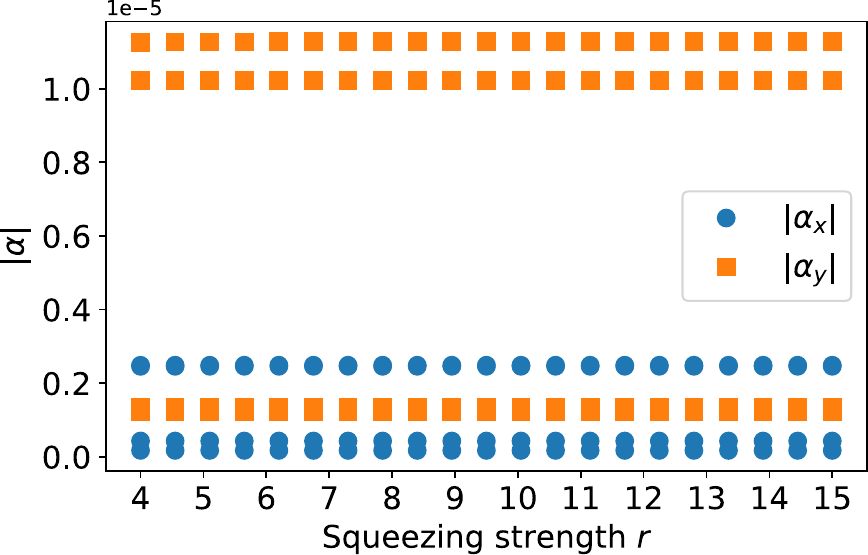}
    \caption{Solutions for $\alpha_x$ and $\alpha_y$ when evaluating Eqs.~\eqref{Eq:SM:SP:ion:QO}-\eqref{Eq:SM:SP:alphay} numerically.~Here, we set $\boldsymbol{p} = \boldsymbol{0}$ (similar results are found for varying values of $\boldsymbol{p}$), $\abs{\boldsymbol{g}(\omega)} = 10^{-8}$ a.u., $\omega = 0.057$ a.u., and $E_{2\omega} = 0.106$ a.u.}
    \label{Fig:Numerics:force}
\end{figure}

These two equations reveal three distinct orders of magnitude for the $\alpha_x$ and $\alpha_y$ variables: terms proportional to $g(\omega)^2$, terms proportional to $g(\omega)$, and terms of order unity.~This motivates a perturbative expansion of $\alpha_x$ and $\alpha_y$ in powers of the coupling parameter $g(\omega)$,
\begin{equation}
	\alpha_i 
	= \alpha_i^{(0)} 
	+ g(\omega)\alpha_i^{(1)}
	+ g(\omega)^2 \alpha_i^{(2)}
	+ \mathcal{O}\big(g(\omega)^3\big), \quad i \in \{x,y\},
\end{equation}
and from Eqs.~\eqref{Eq:SM:SP:alphax:II} and \eqref{Eq:SM:SP:alphay:II}, we find that their zeroth order contributions satisfy
\begin{equation}
	\alpha_y^{(0)} = -i 5 \alpha_x^{(0)},
	\quad
	\alpha_x^{(0)} = -i \alpha_y^{(0)},
\end{equation}
meaning that, at zeroth order with respect to $g(\omega)$, the contribution of an effective photon statistics force vanishes.~We further benchmarked this analytical conclusion against numerical evaluations that do not rely on the Taylor expansions employed above, i.e., solutions of Eqs.~\eqref{Eq:SM:SP:ion:QO}-\eqref{Eq:SM:SP:alphay}. For field amplitudes $\abs{A_{2\omega}(t)} \approx 1$ a.u., we find that $\abs{\alpha_x}\approx \abs{\alpha_y} \propto 10^{-5}$ across values of the squeezing strength ranging from 4 to 15~[Fig.~\ref{Fig:Numerics:force}]. Thus, the perturbative solution considered here appears to capture the correct order of magnitude and provides meaningful insight into the behavior of these variables.

Turning now to the physical implications, these results indicate that, for the present configuration, no photon statistics force arises in the sense described in Refs.~\cite{even_tzur_photon-statistics_2023,rivera-dean_structured_2025}.~The saddle-points, representing the most probable ionization pathways, therefore coincide with those determined primarily by the $2\omega$ field alone. In this picture, the role of the squeezed field $\omega$ is to introduce fluctuations around the ionization trajectories induced by the $2\omega$ field, rather than exerting an effective force that shifts them.

\section{PMD dependence on squeezing parameters}
\label{supp:PMDonr}

\begin{figure}
    \centering
    \includegraphics[width=\linewidth]{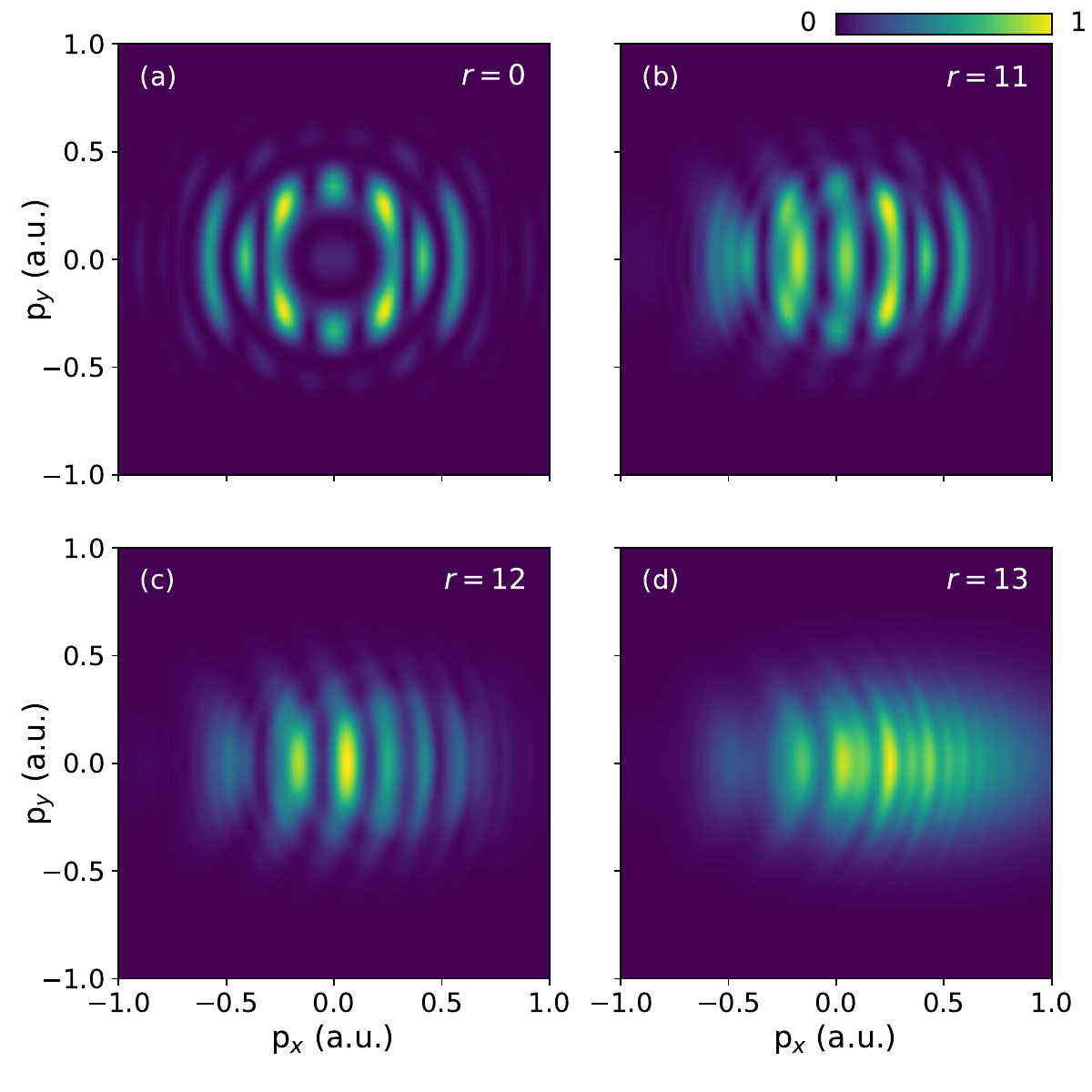}
    \caption{Direct ATI photoelectron momentum distributions with same parameters as in Fig.~\ref{fig:Fig1}(d) for varying squeezing parameter $r$. The squeezing angle is $\phi=0$ for all the plots.}
    \label{fig:PMD_vs_r}
\end{figure}

In this part, we will discuss how the squeezing parameter $r$ and squeezing angle $\phi$ influence the PMDs. We first show that varying the squeezing angle $\phi$ by $\Delta \phi$ is equivalent to shifting the temporal (relative) phase $\theta$ between the two fields by $\Delta \phi/2$. For that, let us look at the Husimi function for BSV [Eq.~\eqref{SM:eq:BSV_Husimi}]. 
Note that $Q_{\text{BSV}}(\alpha;r,\phi+\Delta \phi) = Q_{\text{BSV}}(\alpha e^{-i\Delta\phi/2};r,\phi)$. If we include the temporal phase explicitly in the expression of the electric field, we have (from Eq.~\eqref{eq:Efield}),
\begin{align}
    \boldsymbol{E}_{\boldsymbol{\alpha}}(t)
		&= i\boldsymbol{g}(\omega)
			\big[
				\alpha_\omega e^{-i(\omega t-\theta)} - \alpha_\omega^* e^{i(\omega t-\theta)}
			\big]\nonumber
			\\&\quad
			+ i\boldsymbol{g}(2\omega)
			\big[
				\alpha_{2\omega} e^{-2i\omega t} - \alpha_{2\omega}^* e^{2i\omega t}
			\big].
\end{align}

Thus, $Y_\alpha(\boldsymbol{p})$ depends on $\theta$ through $\boldsymbol{E}_{\boldsymbol{\alpha}}(t)$ and $\boldsymbol{A}_{\boldsymbol{\alpha}}(t)$ (see Eq.~\eqref{SM:eq:P_alpha_SPA}), in particular, $Y_\alpha(\boldsymbol{p},\theta+\Delta\theta) = Y_{\alpha e^{i\Delta \theta}}(\boldsymbol{p},\theta)$. So, the total PMD when the squeezing angle is changed by $\Delta \phi$ is given by 
\begin{align}
  Y(\boldsymbol{p},\theta,\phi&+\Delta\phi)  =\int \dd^2\alpha\, Q_{\text{BSV}}(\alpha;r,\phi+\Delta\phi)
			     Y_{\boldsymbol{\alpha}}(\boldsymbol{p},\theta)\nonumber\\
                 &=\int \dd^2\alpha\, Q_{\text{BSV}}(\alpha e^{-i\Delta\phi/2};r,\phi)
			     Y_{\boldsymbol{\alpha}}(\boldsymbol{p},\theta).
\end{align}
Now, we make the substitution $\beta = \alpha e^{-i\Delta\phi/2}$. The Jacobian satisfies $\dd^2\beta=\dd^2\alpha$. Thus,
\begin{align}
    Y(\boldsymbol{p},\theta,\phi&+\Delta\phi)=\int \dd^2\beta\, Q_{\text{BSV}}(\beta;r,\phi)
			     Y_{\boldsymbol{\beta}e^{i\Delta \phi/2}}(\boldsymbol{p},\theta)\nonumber\\
                 &=\int \dd^2\beta\, Q_{\text{BSV}}(\beta;r,\phi)
			     Y_{\boldsymbol{\beta}}(\boldsymbol{p},\theta+\Delta \phi/2),\\
                 &= Y(\boldsymbol{p},\theta+\Delta \phi/2,\phi).
\end{align}
Hence, a change of squeezing angle by $\Delta \phi$ in the squeezing angle corresponds to a change of $\Delta \phi/2$ in the relative phase between the two superposed fields. 

Let us now look at the effect of the squeezing parameter $r$ on the PMDs. The ATI PMD is calculated from Eq.~\eqref{SM:eq:P_pf_full} after employing the saddle point approximation from Eq.~\eqref{SM:eq:P_alpha_SPA}. Fig.~\ref{fig:PMD_vs_r} shows ATI PMDs for He atom in linearly polarized $\omega-2\omega$ bichromatic light with the coherent $2\omega$ component superposed with the BSV $\omega$ field at different squeezing parameters. The figure shows that the asymmetry, the cutoff and the blurring of the interference patterns increase with squeezing parameter $r$. It starts from a monochromatic field of frequency $2\omega$, with reflection symmetry about the $p_y$ axis, well defined ATI rings and nearly vertical intra-cycle interference fringes [Fig.~\ref{fig:PMD_vs_r}(a)]. For $r=11$ and $r=12$ [Figs.~\ref{fig:PMD_vs_r}(b) and (c), respectively], the symmetry is broken and the features move towards positive momenta $p_x>0$, until the asymmetry surpasses that in the main body of the paper [Fig.~\ref{fig:PMD_vs_r}(d)].

\section{Perturbative expansion of the ATI probability amplitude}\label{Sec:SM:ATI:prob:amp:exp}
In this section, we justify the scaling provided when expanding $b_{\alpha + \delta\alpha}(\boldsymbol{p},t,t_1)$ around $\alpha$.~For simplicity, and given that we are working with linearly polarized fields, we restrict ourselves to a one dimensional analysis.~Then, up to first order in $\delta\alpha$, such a perturbation results in
\begin{equation}
    b_{\alpha + \delta\alpha}(p,t,t_1)
        \approx
            b_{\alpha}(p,t,t_1)
            + \delta \alpha
                \pdv{b_{\alpha}(p,t,t_1)}{\alpha},
\end{equation}
and the derivative with respect to $\alpha$ reads
\begin{equation}
    \begin{aligned}
        &\pdv{b_{\alpha}(p,t,t_1)}{\alpha}
            \\&\hspace{1cm}= i \pdv{S_\alpha(p,t,t_1)}{\alpha}
                e^{iS_\alpha(p,t,t_1)}
                \mel{p+A(t_1)}{\hat{r}}{\text{g}}
                    E_{\alpha}(t_1)
            \\&\hspace{1cm} \quad
                +e^{iS_\alpha(p,t,t_1)}
                \pdv{}{\alpha}
                    \big[
                        \mel{p+A(t_1)}{\hat{r}}{\text{g}}
                    \big]
                    E_{\alpha}(t_1)
            \\&\hspace{1cm} \quad
            +e^{iS_\alpha(p,t,t_1)}
                \mel{p+A(t_1)}{\hat{r}}{\text{g}}
                    \pdv{E_{\alpha}(t_1)}{\alpha}.
    \end{aligned}
\end{equation}
In the following, we evaluate each of the derivatives appearing in the expression independently.

For the first, using the definition of the action, we find
\begin{equation}
    \pdv{S_{\alpha}(p,t,t_1)}{\alpha}
        = - \dfrac{2g(\omega)}{\omega}
            \int^{t}_{t_1}
                \dd \tau
                \big[
                    p+A(\tau)
                \big]
                \cos(\omega \tau + \phi_{\omega}),
\end{equation}
which provides a scaling of the form $\delta E_{\alpha} \sqrt{U_p}/\omega^2$ relative to $b(p,t,t_1)$. For the second term, evaluating the logarithmic derivative of the dipole matrix element to see how it scales with respect to $b(p,t,t_1)$, we find
\begin{equation}
    \begin{aligned}
        &\pdv{}{\alpha}
			\big[
				\ln[\mel{p+A_{\alpha}(t_1)}{\hat{r}}{\text{g}}]
			\big]
			\\&\hspace{1cm}
			= \dfrac{2g(\omega)}{\omega}
			\bigg[
				\dfrac{1}{k(t_1)}
				- \dfrac{6k(t_1)}{k^2(t_1)+\lambda^2}
			\big] \cos(\omega t),
    \end{aligned}
\end{equation}
where $k(t)=p+A(t)$. Given that the dipole matrix element is localized over a momentum range $k(t)\sim \sqrt{I_p}$, this term introduces a relative scaling of $\delta E_{\alpha}/(\omega \sqrt{I_p})$. Finally, for the field derivative term, we obtain
\begin{equation}
    \pdv{E_{\alpha}(t_1)}{\alpha}
        = 2g(\omega)\sin(\omega t_1),
\end{equation}
which scales directly with $\delta E_0$. Collecting all three contributions yields
\begin{equation}
    \begin{aligned}
    &b_{\alpha+\delta\alpha}(\boldsymbol{p},t,t_1)
            \approx b_{\alpha}(\boldsymbol{p},t,t_1)
    \\&\hspace{0.2cm}       
         \times 
         \Bigg[1
                +\delta E_{\alpha}
                \Bigg(
                    \dfrac{1}{E_{\omega,\alpha}+E_{2\omega}} + \dfrac{\sqrt{U_p}}{\omega^2}
                        +\dfrac{1}{\omega\sqrt{I_p}}
                \Bigg)
        \Bigg].
    \end{aligned}
\end{equation}

\end{document}